\documentclass[10pt,letterpaper]{article}

\usepackage[left=2.35cm,right=2.35cm,top=2.4cm,bottom=2.5cm,headheight=22pt,headsep=14pt]{geometry}
\usepackage{titlesec}
\usepackage{fancyhdr}
\usepackage{caption}
\usepackage{enumitem}
\usepackage{mdframed}

\usepackage[utf8]{inputenc} %
\usepackage[T1]{fontenc} %
\usepackage{XCharter} %
\usepackage{natbib} %
\usepackage{hyperref} %
\usepackage{url} %
\usepackage{fontawesome5} %
\usepackage{booktabs} %
\usepackage{longtable} %
\usepackage{amsfonts} %
\usepackage{amsmath}
\usepackage{amssymb}
\usepackage{nicefrac} %
\usepackage{microtype} %
\usepackage{graphicx}
\usepackage{algorithm}
\usepackage{algorithmic}
\usepackage{multirow}
\usepackage[table]{xcolor}

\definecolor{HarnessInk}{HTML}{173846}
\definecolor{HarnessTeal}{HTML}{0E7C7B}
\definecolor{HarnessMint}{HTML}{E8F4F2}
\definecolor{HarnessAmber}{HTML}{D99635}
\definecolor{HarnessSlate}{HTML}{5F7078}
\definecolor{HarnessPale}{HTML}{F5F8F8}
\definecolor{DesignerBlue}{HTML}{2477B5}
\definecolor{CriticizerGreen}{HTML}{3D852B}
\definecolor{CascadeNavy}{HTML}{14566B}
\definecolor{AnalyzerOrange}{HTML}{A94E1B}

\hypersetup{
  colorlinks=true,
  linkcolor=HarnessTeal,
  citecolor=HarnessTeal,
  urlcolor=HarnessInk,
  pdftitle={EvoSafeHarness: Evolving Model- and Domain-Specific Harnesses for Securing Agents},
  pdfsubject={Automated synthesis of deployment-specific safety harnesses for tool-using agents}
}

\setlist[itemize]{leftmargin=1.4em,itemsep=1pt,topsep=3pt,parsep=0pt}
\setlist[enumerate]{leftmargin=1.6em,itemsep=1pt,topsep=3pt,parsep=0pt}

\titleformat{\section}
  {\Large\bfseries\color{HarnessInk}}
  {\colorbox{HarnessMint}{\color{HarnessTeal}\sffamily\bfseries\thesection}}
  {0.65em}{}
\titleformat{\subsection}
  {\large\bfseries\color{HarnessTeal}}
  {\thesubsection}{0.55em}{}
\titleformat{\subsubsection}
  {\normalsize\bfseries\color{HarnessInk}}
  {\thesubsubsection}{0.5em}{}
\titleformat{\paragraph}[runin]
  {\bfseries\color{HarnessInk}}{}{0pt}{}[.]
\titlespacing*{\section}{0pt}{3.2ex plus .8ex minus .3ex}{1.0ex}
\titlespacing*{\subsection}{0pt}{2.3ex plus .5ex minus .2ex}{0.65ex}
\titlespacing*{\subsubsection}{0pt}{1.8ex plus .4ex minus .2ex}{0.45ex}
\titlespacing*{\paragraph}{0pt}{1.2ex}{0.55em}

\renewcommand{\headrulewidth}{0.8pt}
\renewcommand{\headrule}{%
  \hbox to\headwidth{\color{HarnessTeal}\leaders\hrule height \headrulewidth\hfill}}
\renewcommand{\footrulewidth}{0pt}

\fancypagestyle{modernfirst}{
  \fancyhf{}
  \fancyfoot[L]{\sffamily\footnotesize\color{HarnessSlate}EvoSafeHarness}
  \fancyfoot[R]{\sffamily\footnotesize\color{HarnessSlate}arXiv preprint \textbullet\ \thepage}
  \renewcommand{\headrulewidth}{0pt}
  \renewcommand{\footrulewidth}{0.8pt}
  \renewcommand{\footrule}{%
    \hbox to\headwidth{\color{HarnessTeal}\leaders\hrule height \footrulewidth\hfill}}
}

\makeatletter
\renewcommand{\maketitle}{%
  \thispagestyle{modernfirst}%
  \begingroup
  \vspace*{-2.05cm}
  {\centering\fontsize{19}{23}\selectfont\bfseries\color{HarnessInk}\@title\par}
  \vspace{0.55em}
  {\color{HarnessTeal}\rule{\textwidth}{1.4pt}\par}
  \vspace{0.45em}
  {\centering\color{HarnessInk}\@author\par}
  \vspace{2.0em}
  \endgroup
}
\makeatother

\renewenvironment{abstract}
  {\par\medskip
   \begin{mdframed}[
     backgroundcolor=HarnessMint,
     linecolor=HarnessTeal,
     leftline=true,
     rightline=false,
     topline=false,
     bottomline=false,
     linewidth=3pt,
     innerleftmargin=10pt,
     innerrightmargin=10pt,
     innertopmargin=6pt,
     innerbottommargin=6pt,
     skipabove=26pt,
     skipbelow=12pt]
   \small\noindent{\sffamily\bfseries\color{HarnessTeal}ABSTRACT}\par\smallskip}
  {\end{mdframed}\par\medskip}

\renewcommand{\arraystretch}{1.06}

\definecolor{ourshl}{HTML}{E6F4EA}
\definecolor{PendingRed}{HTML}{B42318}

\definecolor{lessonbg}{HTML}{E8F4F2}
\definecolor{lessonrule}{HTML}{0E7C7B}

\title{EvoSafeHarness: Evolving Model- and Domain-Specific\\Harnesses for Securing Agents}

\author{%
  {\normalsize\bfseries
   Nanxi Li\textsuperscript{1}\quad
   Yingzi Ma\textsuperscript{2}\quad
   Yulong Cao\textsuperscript{3}\quad
   Chong Xiang\textsuperscript{3}\quad
   G. Edward Suh\textsuperscript{3}\\[1pt]
   Bo Li\textsuperscript{4}\quad
   Yejin Choi\textsuperscript{5}\quad
   Dawn Song\textsuperscript{6}\quad
   Chaowei Xiao\textsuperscript{1,3}}\\[3pt]
  {\small\color{HarnessSlate}
   \textsuperscript{1}Johns Hopkins University\quad
   \textsuperscript{2}University of Wisconsin--Madison\quad
   \textsuperscript{3}NVIDIA\\
   \textsuperscript{4}University of Illinois Urbana--Champaign\quad
   \textsuperscript{5}Stanford University\quad
   \textsuperscript{6}University of California, Berkeley}%
}

\begin{document}
\maketitle

\begingroup
\centering
\makeatletter\def\@captype{figure}\makeatother
\includegraphics[width=\textwidth]{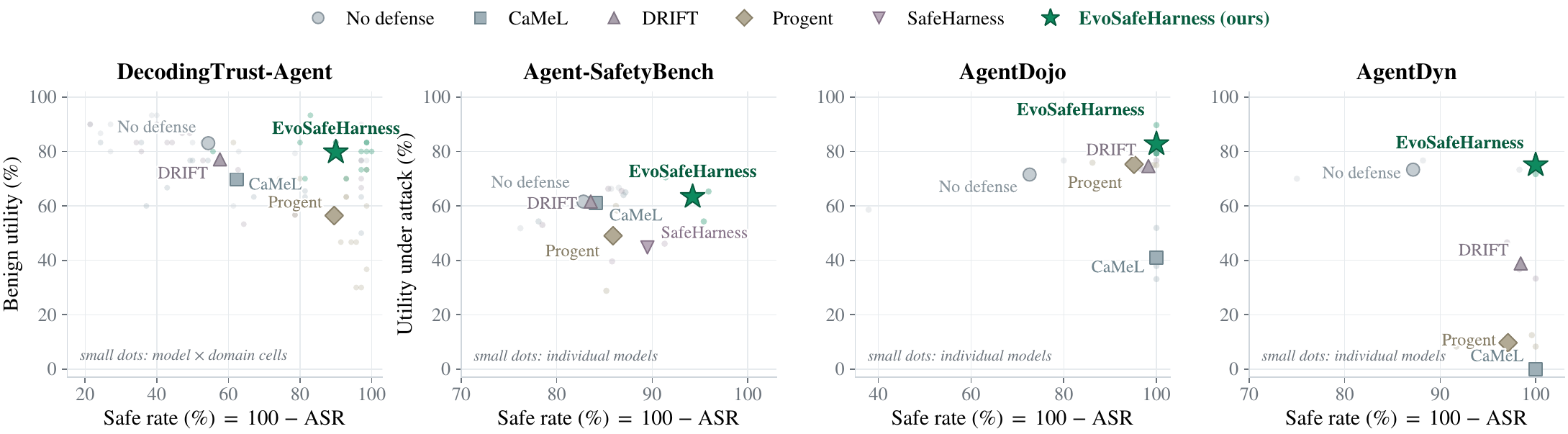}
\caption{\textbf{EvoSafeHarness dominates the utility--security trade-off on all
four benchmarks.}
Each panel plots utility against safe rate ($100-\mathrm{ASR}$), so higher is better on
both axes and the ideal defense sits in the top-right corner. Large markers show
cross-model method averages and small dots show individual evaluation cells
(model$\times$domain cells on DecodingTrust-Agent, victim models elsewhere). The
Agent-SafetyBench panel reports utility under attack; the AgentDojo panel reports held-out
test results and the AgentDyn panel zero-shot transfer of the AgentDojo harness.
}
\label{fig:headline-results}
\endgroup

\begin{abstract}
Large Language Model (LLM) agents turn language into real-world effects and must stay safe against both indirect prompt injections and direct harmful requests. 
System-level safety harnesses add an enforcement layer around the model, but existing harnesses are
built once by experts and applied across heterogeneous models and domains. The effective defense is
deployment-dependent: models differ in how much external enforcement they tolerate before utility
drops, and domains differ in which effects, state, and action sequences must be governed. A harness
strict enough for one model over-blocks another, and a policy general enough to transfer across
domains misses the safety relations of the application.

We present \textbf{\textsc{EvoSafeHarness}}, a safety-specific harness optimization framework that
synthesizes a deployable harness for a frozen model in a target domain. Unlike harness generation
frameworks that target utility alone, it jointly searches a natural-language policy and executable
code logic, guided by the target model's behavioral feedback and a domain specification, and screens
each candidate with a fresh-context adversarial review that rejects rules keyed to benchmark
artifacts.
Across four agent benchmark families, \textsc{EvoSafeHarness} establishes a stronger safety--utility
frontier than fixed expert-designed defenses. On DecodingTrust-Agent, across fifteen independently
searched model$\times$domain deployments, it reduces attack success rate (ASR) from $50.9\%$ to
$12.6\%$ for direct attacks and from $40.4\%$ to $7.4\%$ for indirect attacks at a $3.3$-point
utility cost. On
AgentDojo it reaches $82.8\%$ utility at $0.0\%$ ASR, twice the utility of CaMeL at the same
zero-ASR operating point, and the same harness transfers unchanged to unseen AgentDyn suites. It
also attains the best score on Agent-SafetyBench for every victim and keeps ASR below $18\%$
against an adaptive PAIR-style attacker with $16$ attempts per task.
Analysis of the synthesized harnesses shows that domain semantics determine which safety relations
and trajectory state are required, while model and runtime behavior determine how and where those
relations are enforced.

\end{abstract}

\par\vspace{10pt}
{\centering\small
\href{https://github.com/SaFo-Lab/EvoSafeHarness}{\faGithub~\texttt{github.com/SaFo-Lab/EvoSafeHarness}}\qquad
\href{https://andylinx.github.io/EvoSafeHarness/}{\faGlobe~\texttt{andylinx.github.io/EvoSafeHarness}}\par}
\vspace{3pt}

\clearpage

\section{Introduction}

Language-model agents are moving from demonstration to deployment. Agent safety becomes more important and urgent.
What makes agent safety qualitatively harder is that harmful instructions can enter
through two channels that cross different security boundaries. In an \emph{indirect
prompt injection} attack \citep{greshake,liu2024formalizing,perez}, an adversary embeds
instructions in external content---such as an email, a web page, or a document---that the
agent must consume as data. If the agent treats this content as authoritative, it may
execute actions that the user never requested. In a \emph{direct attack}
\citep{dtap,agentharm,agentsafetybench}, the harmful instruction arrives through the user
channel itself: a malicious user asks the agent to transfer funds beyond an
approved limit, delete production data, or disclose credentials. 

\begin{figure*}[t!]
\centering
\includegraphics[width=\textwidth]{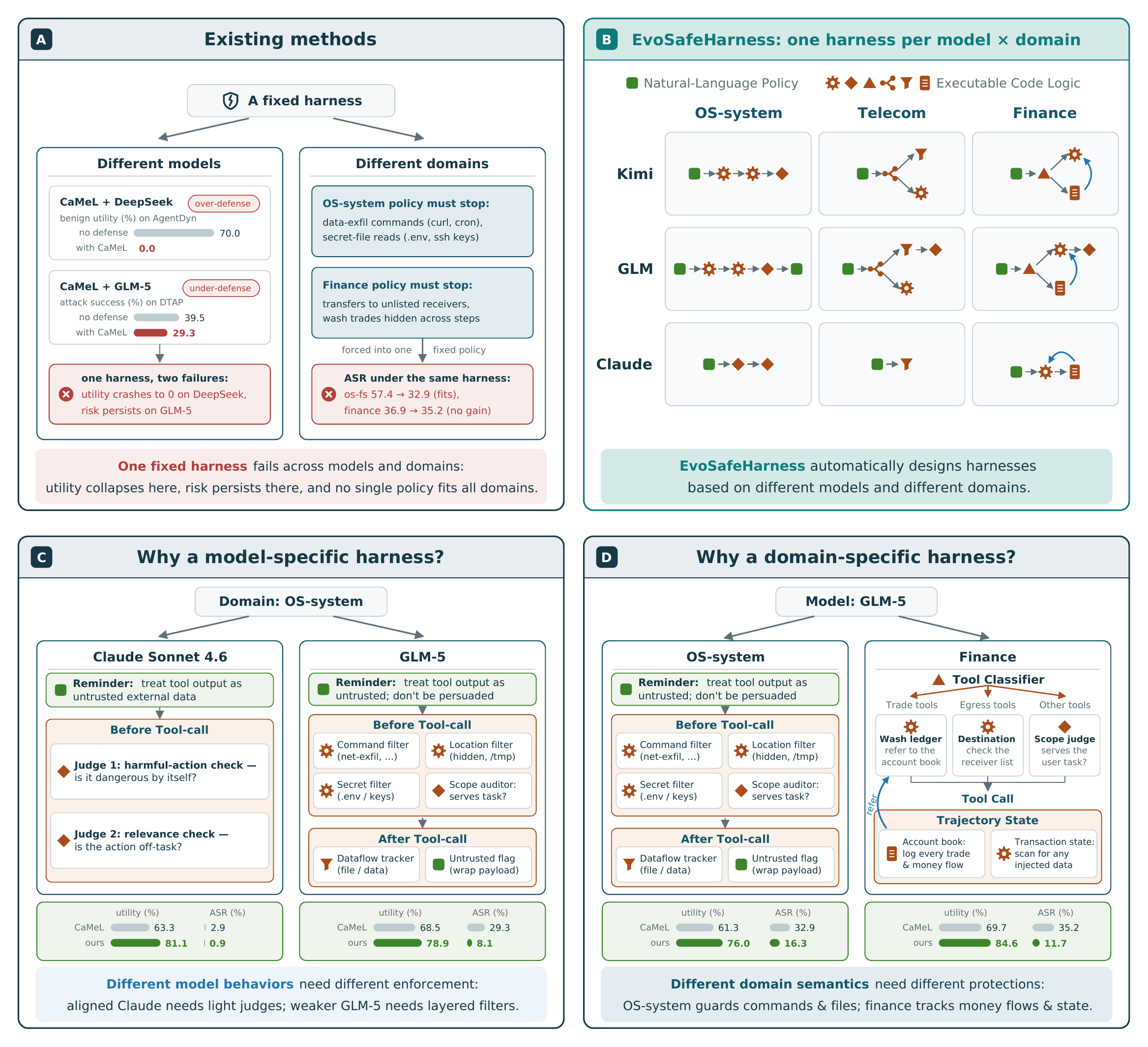}
\caption{\textbf{From the limits of fixed defenses to model- and domain-specific \textsc{EvoSafeHarness}.}
(A) A fixed harness fails across models and domains: utility collapses on one victim, risk persists
on another, and no single policy fits every domain. (B) \textsc{EvoSafeHarness} designs a distinct combination
of a natural-language policy bundle $P$ and an executable runtime enforcement program $C$ per model$\times$domain deployment. (C) With
the domain fixed, model behavior determines how the same relation is enforced. (D) With the model
fixed, domain semantics determine which relations and state must be protected. In both, each
searched harness beats CaMeL on utility and ASR.}
\label{fig:intuition}
\end{figure*}

To address these risks, model-level defenses \citep{instructionhierarchy,struq,secalign} train the model to separate
trusted instructions from untrusted content or to refuse unsafe requests. They make compliant
behavior more likely but provide no deterministic enforcement boundary independent of the model. 
This limitation has motivated system-level defenses implemented at the
\emph{harness} level. Existing approaches \citep{spotlighting,llamaguard,drift,camel,safeharness} use
mechanisms such as provenance marking, injection detection, trajectory validation, capability
enforcement, and lifecycle-level execution control, demonstrating that the
harness is an effective locus of agent security. Each of them, however, is a fixed,
expert-designed harness whose policy and control flow are the same across heterogeneous models and
domains (Figure~\ref{fig:intuition}A). This leaves an open question: can a universal harness
provide the best safety--utility trade-off across deployments, or should the harness be adapted to the target model and domain?

Building a universal harness that remains both safe and useful across models and
applications is challenging because these two sources of variation affect different
aspects of harness design. As shown in
Figure~\ref{fig:intuition},
model variation changes
the appropriate strength of enforcement. 
A strongly safety-trained model already resists many attacks  that a harness is meant to block, so a strict capability-based defense such as CaMeL \citep{camel}
buys little security improvement but at a large utility cost, while the same restrictions may be essential for a weaker one. Application domain variation goes beyond enforcement strength: it changes the safety relations and control flow that
the harness must implement. This is particularly clear for direct attacks, where the
request arrives through the user channel and its harmfulness must be determined from the
semantics of the requested effect. A filesystem harness may need to inspect command effects, sensitive paths, and secret movement while a
finance harness must separate trades from money egress and keep transaction history, since
permissible trades can compose into wash trading. Thus,  a single universal harness is unlikely to provide
the best safety--utility trade-off across both.

This heterogeneity makes deployment-specific harness generation an appealing next step. In general domains, automatic harness optimization~\citep{metaharness,adas,aflow,ahe} has made significant progress by designing a proposer to iteratively generate harness code
based on the target model's behavioral feedback.
Similarly, the same behavioral signals can expose model-specific security failures. However, simply replacing
the task feedback with attack success introduces three challenges.

First,  a harness can appear safe by refusing everything, converting security into utility collapse. 
Second, since evaluation data directly guide the optimizer, data-driven optimization can exploit superficial artifacts, such as
filenames, paths, or recurring attack phrases, and thereby produce improvements that vanish under trivial renaming or paraphrasing.
Third,   security evaluation requires broad coverage on diverse direct and indirect attacks.  Yet executing every candidate on the full set of long-horizon agent trajectories is prohibitively expensive, particularly when search is repeated across model--domain deployments.
Safety harness synthesis therefore needs a search that preserves utility, tests generalization, and keeps the whole optimization efficient. 

\begin{figure*}[t]
\centering
\includegraphics[width=\textwidth]{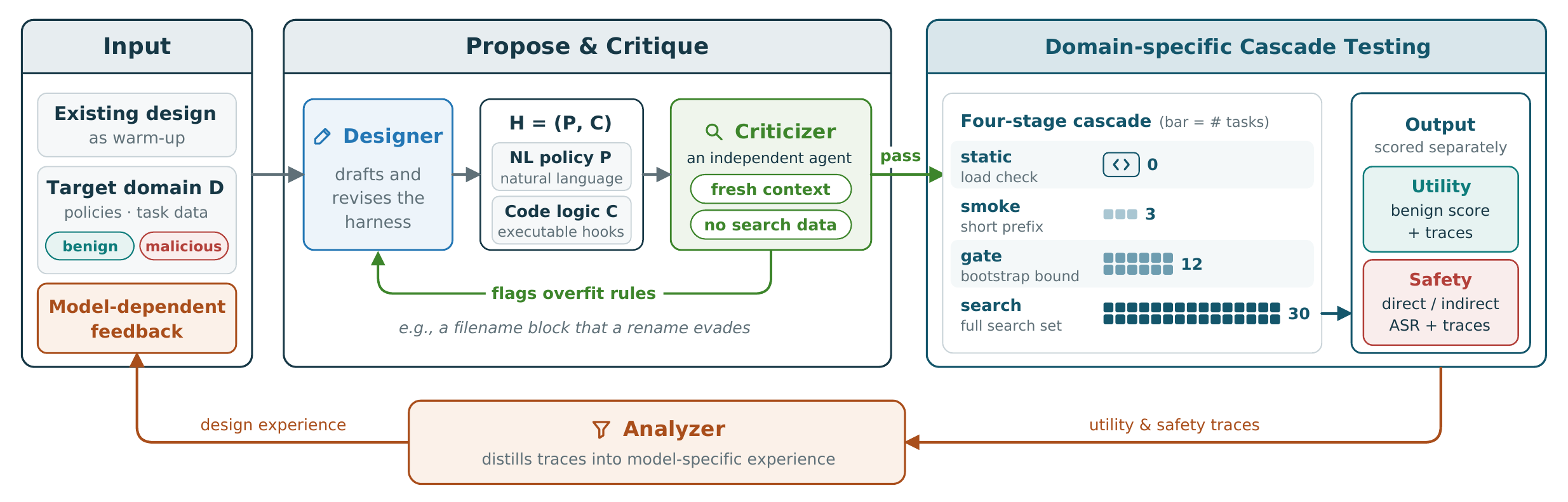}
\caption{\textbf{\textsc{EvoSafeHarness} searches a deployment-specific security harness.} The
\textcolor{DesignerBlue}{Designer} drafts a harness $H=(P,C)$ from existing defenses, the target
domain $D$, and model-dependent feedback. An independent \textcolor{CriticizerGreen}{Criticizer} that
never sees search data sends overfit rules back for revision. Harnesses that pass go through a
four-stage \textcolor{CascadeNavy}{cascade} that scores utility and safety separately, and the
\textcolor{AnalyzerOrange}{Analyzer} turns the resulting traces into feedback.}
\label{fig:overview}
\end{figure*}

We present \textbf{\textsc{EvoSafeHarness}}, a safety-specific harness optimization framework that searches a deployable
harness $H$ around a frozen model $M$ in a target domain $D$ (Figure~\ref{fig:overview}). 
Each search is governed by a domain specification that defines the deployment environment and security requirements. Starting from
existing harness designs as a warm start, our \textcolor{DesignerBlue}{\textbf{Designer}} iteratively generates a harness, consisting of
both a natural-language policy bundle $P$ and an executable runtime enforcement program $C$, from the archived outcomes and failure traces of the target model.
Each proposed harness then passes through two safety-specific stages before its feedback can guide
the next iteration. First, a data-independent
\textcolor{CriticizerGreen}{\textbf{Criticizer}} inspects the proposed policy and code without accessing the evaluation instances,  preventing shortcut learning. Candidates whose rules depend on superficial evaluation artifacts are identified, addressing the second challenge.
Candidates that pass this review enter a four-stage
\textcolor{CascadeNavy}{\textbf{Cascade Test Environment}} consisting of static validation, smoke testing, a
confidence-based gate, and evaluation on the full search set. The cascade reuses evidence across stages and stops clearly inferior candidates early, addressing the third challenge.
The Cascade measures benign utility, direct-attack success, and indirect-attack success separately.
An \textcolor{AnalyzerOrange}{\textbf{Analyzer}} converts these outcomes and their trajectories into
structured failure patterns and design lessons, which are added to the archive for the next
Designer iteration. Because the Designer receives evidence from both benign and adversarial tasks,
it can improve security without treating refusal or utility loss as progress, addressing the first challenge. 

We evaluate \textsc{EvoSafeHarness} on four agent benchmark families spanning both attack channels,
and our method achieves a stronger safety--utility trade-off on all of them compared to existing methods. On DecodingTrust-Agent, our optimized harnesses cut direct ASR from
$50.9\%$ to $12.6\%$ and indirect ASR from $40.4\%$ to $7.4\%$ at a $3.3$-point utility cost on average,
whereas the fixed defenses CaMeL and DRIFT leave direct ASR above $44\%$ and indirect ASR above $31\%$,
and Progent sacrifices more than $20$ utility points
(Table~\ref{tab:dt-crossmodel}). On AgentDojo, our harness reaches $82.8\%$ utility at $0.0\%$ ASR, twice the
utility of CaMeL at the same zero ASR. The same harness also transfers zero-shot to
unseen AgentDyn suites, evidence that it captures general safety designs rather than artifacts of the search tasks.
Against adaptive attacks, a frozen Agent-SafetyBench harness keeps ASR below $18\%$ even when a PAIR-style attacker~\citep{pair} gets $16$ attempts per task.

Beyond aggregate results, the generated harnesses show why model- and domain-specific design produces these gains (Figure~\ref{fig:intuition}C--D): With the domain fixed, a safer Sonnet~4.6 reaches zero held-out ASR with a light policy and two semantic checks, whereas a less safe GLM-5 needs more deterministic gates and provenance
state. With the model fixed, the harness protects command effects and paths in the filesystem, but
trade-versus-egress distinctions and transaction history in finance.

\section{Method: \textsc{EvoSafeHarness}}
\label{sec:method}

\paragraph{Threat Models}\label{sec:threat-model}

A tool-using agent is a pair $(M, H)$: a frozen language model $M$ and a \emph{harness} $H$ that mediates every interaction between $M$, the user, and the tools.  We consider the commonly used threat model~\citep{agentdojo,dtap,agentsafetybench,agentdyn} with two attack channels. In an \emph{indirect attack}, the attacker controls external
content that the agent must consume as data, such as a file, message, record, or tool result, but
does not control the user's request. The attack succeeds if this content redirects the agent toward
an unauthorized goal. In a \emph{direct attack}, the harmful goal is supplied through the user channel itself. In both cases, the attacker may adapt the malicious input but cannot modify the model weights, harness code, trusted deployment policy, or evaluation judges. 
Here, as agent developers, we consider the practical setting where we cannot update model parameters, and instead design the harness against both above threats. 

\paragraph{What Should a Security Harness Consist of?}
Under this threat model, securing an agent requires more than detecting malicious strings. Whether an action is safe depends on semantic relationships among its
inputs and effects: which source supplied an instruction, whether that source is
trusted, whether the proposed action serves the user's intent, and whether its
effects are authorized by the domain policy. A security harness must therefore
interpret concepts such as trust, provenance, task relevance, and authorization.
Semantic relationship alone, however, does not provide reliable enforcement. If the
target model is responsible for both interpreting a security rule and choosing
whether to follow it, an attack can compromise the component due to the probabilistic nature of the model-level enforcement. The harness must also control the model-external execution path: what information each component receives, which security checks run and when, what state persists across steps, and whether a proposed tool call is allowed to execute.
These requirements motivate a harness design space that jointly captures
\emph{semantic security specification} and \emph{runtime enforcement}.

Therefore, we represent a harness as a pair $H = (P, C)$,
where $P$ is a bundle of natural-language security policies and $C$ is an executable runtime enforcement program. The policy bundle $P$ specifies the security semantics used by every model in the harness, including the system policy of the target model
and the policies assigned to auxiliary guard models. These policies define how the
models should reason about trust boundaries, provenance, user intent, task relevance,
and authorization. 

The executable component $C$ realizes the harness's model-external control and data flow. It determines which model, checker, or tool is invoked at each point; what information each component receives; how outputs are transformed and routed; what state is maintained across the trajectory; and how the resulting security decision is enforced. For example, $C$ may apply a deterministic check to a proposed tool
call, invoke a guard model when semantic judgment is required, combine the resulting
signals, and finally allow, rewrite, or block the call. Although $C$ may invoke
probabilistic model-based judges, their invocation, isolation, input construction, and verdict enforcement are controlled by executable logic outside the target model.
Both components are necessary. Optimizing $P$ alone leaves enforcement probabilistic, since it depends only on model compliance, and cannot introduce system-level deterministic enforcement. Optimizing only a fixed enforcement program, in contrast, restricts the harness to predefined security concepts and decision rules.

\subsection{Feedback-Guided Optimization Pipeline}
\label{sec:pipeline}

Figure~\ref{fig:overview} presents the complete optimization pipeline, and Algorithm~\ref{alg:loop} in the appendix gives its pseudocode. Each search
is guided by (1) a domain specification $\Sigma_d$, which describes the target model,
tools, trust boundaries, security requirements, evaluator interfaces, and immutable
application constraints  as detailed in Appendix~\ref{app:impl}. It specifies what behavior must be protected without
prescribing which defense to build; and (2) an archive $\mathcal{R}$ containing previously proposed harnesses, their evaluation outcomes evaluated by the agent with provided data, and the corresponding failure traces. At the first search round, the archive is
initialized with the no-defense harness and security principles distilled from
existing defenses~\citep{drift,camel}. These seeds provide a warm start but do not constrain the
resulting architecture.

At iteration $t$, the \textcolor{DesignerBlue}{\textbf{Designer}} reads the domain specification and accumulated feedback and proposes
$H_t=(P_t,C_t)=\operatorname{Designer}(\Sigma_d,\mathcal{R}_{t-1})$.
The revision is free-form: the Designer may revise the policy bundle $P$, or rewrite the enforcement program $C$ entirely. 
Before evaluation, an independent \textcolor{CriticizerGreen}{\textbf{Criticizer}} inspects $H_t$ for rules that depend on
superficial artifacts of the evaluated data. Candidates that pass this review enter
the \textcolor{CascadeNavy}{\textbf{Cascade Test Environment}}, which evaluates them on increasingly broad subsets
of the search set. The evaluation returns both
aggregate outcomes and the corresponding execution trajectories.
The \textcolor{AnalyzerOrange}{\textbf{Analyzer}}  converts this evidence into two persistent forms of feedback. A
failure-pattern log groups successful attacks by attack channel and violated
security invariant, and benign failures by their cause, while a design-lesson log records which policy and enforcement
mechanisms succeeded or failed. These summaries, together with the raw trajectories
and candidate scores, are added to $\mathcal{R}_t$ and guide the next Designer
iteration.
After a fixed search budget, \textsc{EvoSafeHarness} returns the highest-scoring candidate (by attack success rate and benign utility)
that completes the full evaluation. Held-out tasks are used only for final evaluation.

\subsection{Security-Specific Optimization Designs}

A generic feedback-driven harness optimizer is insufficient in this setting.
Security introduces three problems: attack-only feedback rewards refusal, repeated
evaluation feedback encourages artifact-specific shortcuts, and full security
evaluation is too expensive to run for every candidate. \textsc{EvoSafeHarness} addresses
these problems through the following designs, ablated in \S\ref{sec:results-ablations}.

 \noindent\textbf{Balanced behavioral feedback.}
Different from harness optimization in the general domain, security optimization cannot be guided by attack success alone, since a harness can then appear safe by refusing everything, trading security for utility collapse. 
Moreover, an aggregate score cannot reveal whether a failure arises
from incorrect authorization reasoning, a broken trust boundary, or missing runtime
enforcement.
\textsc{EvoSafeHarness} therefore treats evaluation as a source of behavioral evidence. For each candidate, the evaluation
environment preserves benign, direct-attack, and indirect-attack trajectories
separately. The \textcolor{AnalyzerOrange}{\textbf{Analyzer}}  interprets these trajectories jointly with the candidate's
policy and enforcement logic. It maintains a persistent failure-pattern log that
organizes failures by attack channel and violated security invariant, and a
design-lesson log that records which changes to $P$ or $C$ improved security, caused
utility regressions, or failed to address the underlying vulnerability.

 \noindent\textbf{Evaluation-instance-independent critique.}
Because evaluation feedback directly guides later proposals, the Designer may learn
rules that exploit recurring filenames, paths, tool arguments, or attack phrases.
Such a candidate may perform well on the search set while failing under a simple
rename or paraphrase. Running more evaluations on the same data does not expose this
failure because the artifact remains shared across those evaluations.

The \textcolor{CriticizerGreen}{\textbf{Criticizer}}  therefore examines the candidate independently of the evaluation
instances. It receives $H_t$ and $\Sigma_d$, but not the search tasks, earlier
outcomes, or the Designer's reasoning. For each rule, it considers inexpensive
intent-preserving evasions, such as renaming a file, relocating content, paraphrasing
an instruction, or replacing a tool call with an equivalent action. It flags rules
whose effectiveness depends on literal artifacts not justified by $\Sigma_d$ and
returns them to the Designer for revision. Because this review occurs before
evaluation, an artifact-specific rule cannot earn a high score and propagate through
the archive unchecked.

 \noindent\textbf{Cascaded evaluation.}
Reliable security feedback requires broad benign and adversarial coverage, but
executing the complete task set for every candidate is prohibitively expensive.
Evaluating every candidate on a small random subset would be cheaper, but could miss
entire attack surfaces. \textsc{EvoSafeHarness} instead uses four nested stages in its \textcolor{CascadeNavy}{\textbf{Cascade Test Environment}}:
(1)~\textbf{Static validation} rejects candidates that fail to load, violate
the harness interface, or modify immutable components.
(2)~\textbf{Smoke testing} runs a small mixed prefix of benign and adversarial
tasks to detect broken execution, severe over-refusal, and clearly ineffective
defenses.
(3)~\textbf{Confidence-based gating} compares the accumulated evidence with
the no-defense reference. A candidate is stopped only when a one-sided bootstrap
upper bound places it confidently below the reference by a predefined margin.
(4)~\textbf{Full evaluation} runs the surviving candidate on the complete
search set and reports utility, direct ASR, and indirect ASR separately.
These stages form a single nested evidence stream: each stage extends the
trajectories collected by the previous stage rather than restarting evaluation.
The cascade therefore removes clearly inferior candidates early while preserving
complete benign and adversarial coverage for candidates that remain competitive.

\begin{table}[t]
\centering
\small
\setlength{\tabcolsep}{3.6pt}
\renewcommand{\arraystretch}{1.0}
\begin{tabular}{@{}l@{\hspace{3pt}}l ccc ccc ccc ccc@{}}
\toprule
\multirow{2}{*}{\textbf{Model}} & \multirow{2}{*}{\textbf{Defense}}
 & \multicolumn{3}{c}{\textbf{os-fs}} & \multicolumn{3}{c}{\textbf{finance}} & \multicolumn{3}{c}{\textbf{telecom}} & \multicolumn{3}{c}{\textbf{Average}} \\
\cmidrule(lr){3-5}\cmidrule(lr){6-8}\cmidrule(lr){9-11}\cmidrule(lr){12-14}
 & & Util & D-ASR & I-ASR & Util & D-ASR & I-ASR & Util & D-ASR & I-ASR & Util & D-ASR & I-ASR \\
\midrule
\multirow{5}{*}{Sonnet 4.6}
 & No-Defense & 86.7 & 5.7 & 14.3 & 90.0 & 0.0 & 5.7 & 76.7 & 0.0 & 2.9 & 84.5 & 1.9 & 7.6 \\
 & CaMeL & 50.0 & 5.7 & 0.0 & 63.3 & 0.0 & 5.7 & 76.7 & 0.0 & 5.7 & 63.3 & 1.9 & 3.8 \\
 & DRIFT & 66.7 & 5.7 & 0.0 & 70.0 & 0.0 & 5.7 & 76.7 & 0.0 & 5.7 & 71.1 & 1.9 & 3.8 \\
 & Progent & 36.7 & 2.9 & 0.0 & 60.0 & 0.0 & 2.9 & 73.3 & 2.9 & 0.0 & 56.7 & 1.9 & 1.0 \\
 \rowcolor{ourshl}
 & EvoSafeHarness & 80.0 & 0.0 & 0.0 & 80.0 & 0.0 & 2.9 & 83.3 & 2.9 & 0.0 & 81.1 & 1.0 & 1.0 \\
\midrule
\multirow{5}{*}{GLM-5}
 & No-Defense & 83.3 & 57.1 & 42.9 & 79.3 & 17.1 & 25.7 & 83.3 & 62.9 & 31.4 & 82.0 & 45.7 & 33.3 \\
 & CaMeL & 63.3 & 20.0 & 17.1 & 65.5 & 11.4 & 25.7 & 76.7 & 61.8 & 40.0 & 68.5 & 31.1 & 27.6 \\
 & DRIFT & 73.3 & 50.0 & 24.2 & 56.7 & 11.4 & 31.4 & 86.7 & 55.9 & 44.0 & 72.2 & 39.1 & 33.2 \\
 & Progent & 70.0 & 8.6 & 5.7 & 46.7 & 2.9 & 5.7 & 83.3 & 37.1 & 40.0 & 66.7 & 16.2 & 17.1 \\
 \rowcolor{ourshl}
 & EvoSafeHarness & 80.0 & 17.1 & 5.7 & 83.3 & 8.6 & 11.4 & 73.3 & 5.7 & 0.0 & 78.9 & 10.5 & 5.7 \\
\midrule
\multirow{5}{*}{Kimi-K2.5}
 & No-Defense & 80.0 & 94.3 & 51.4 & 60.0 & 42.9 & 34.3 & 73.3 & 77.1 & 31.4 & 71.1 & 71.4 & 39.0 \\
 & CaMeL & 66.7 & 25.7 & 20.0 & 63.3 & 34.3 & 31.4 & 76.7 & 77.1 & 22.9 & 68.9 & 45.7 & 24.8 \\
 & DRIFT & 83.3 & 91.4 & 20.0 & 53.3 & 34.3 & 37.1 & 86.7 & 74.3 & 31.4 & 74.4 & 66.7 & 29.5 \\
 & Progent & 30.0 & 8.6 & 0.0 & 46.7 & 8.6 & 8.6 & 56.7 & 40.0 & 2.9 & 44.4 & 19.0 & 3.8 \\
 \rowcolor{ourshl}
 & EvoSafeHarness & 70.0 & 31.4 & 5.7 & 83.3 & 8.6 & 11.4 & 80.0 & 2.9 & 2.9 & 77.8 & 14.3 & 6.7 \\
\midrule
\multirow{5}{*}{Qwen3.7-plus}
 & No-Defense & 86.7 & 77.1 & 74.3 & 93.3 & 57.1 & 62.9 & 80.0 & 28.6 & 45.7 & 86.7 & 54.3 & 61.0 \\
 & CaMeL & 60.0 & 82.9 & 42.9 & 73.3 & 51.4 & 62.9 & 76.7 & 48.6 & 42.9 & 70.0 & 61.0 & 49.6 \\
 & DRIFT & 83.3 & 77.1 & 54.3 & 83.3 & 51.4 & 62.9 & 76.7 & 42.9 & 51.4 & 81.1 & 57.1 & 56.2 \\
 & Progent & 46.7 & 8.6 & 2.9 & 63.3 & 5.7 & 8.6 & 70.0 & 37.1 & 2.9 & 60.0 & 17.1 & 4.8 \\
 \rowcolor{ourshl}
 & EvoSafeHarness & 70.0 & 0.0 & 14.3 & 93.3 & 11.4 & 22.9 & 73.3 & 2.9 & 0.0 & 78.9 & 4.8 & 12.4 \\
\midrule
\multirow{5}{*}{\shortstack[l]{DeepSeek\\V4-Flash}}
 & No-Defense & 90.0 & 94.3 & 62.9 & 93.3 & 65.7 & 57.1 & 90.0 & 82.9 & 62.9 & 91.1 & 81.0 & 61.0 \\
 & CaMeL & 66.7 & 91.4 & 22.9 & 83.3 & 68.6 & 60.0 & 83.3 & 85.7 & 65.7 & 77.8 & 81.9 & 49.5 \\
 & DRIFT & 90.0 & 97.1 & 60.0 & 90.0 & 60.0 & 60.0 & 80.0 & 88.6 & 40.0 & 86.7 & 81.9 & 53.3 \\
 & Progent & 30.0 & 2.9 & 2.9 & 63.3 & 5.7 & 8.6 & 70.0 & 48.6 & 2.9 & 54.4 & 19.0 & 4.8 \\
 \rowcolor{ourshl}
 & EvoSafeHarness & 80.0 & 74.3 & 14.3 & 83.3 & 20.0 & 20.0 & 83.3 & 2.9 & 0.0 & 82.2 & 32.4 & 11.4 \\
\midrule
\multirow{5}{*}{\textbf{Average}}
 & No-Defense & 85.3 & 65.7 & 49.1 & 83.2 & 36.6 & 37.1 & 80.7 & 50.3 & 34.9 & 83.1 & 50.9 & 40.4 \\
 & CaMeL & 61.3 & 45.1 & 20.6 & 69.7 & 33.1 & 37.1 & 78.0 & 54.6 & 35.4 & 69.7 & 44.3 & 31.1 \\
 & DRIFT & 79.3 & 64.3 & 31.7 & 70.7 & 31.4 & 39.4 & 81.4 & 52.3 & 34.5 & 77.1 & 49.3 & 35.2 \\
 & Progent & 42.7 & 6.3 & 2.3 & 56.0 & 4.6 & 6.9 & 70.7 & 33.1 & 9.7 & 56.4 & 14.7 & 6.3 \\
 \rowcolor{ourshl}
 & EvoSafeHarness & 76.0 & 24.6 & 8.0 & 84.6 & 9.7 & 13.7 & 78.6 & 3.4 & 0.6 & 79.8 & 12.6 & 7.4 \\
\bottomrule
\end{tabular}
\caption{\textbf{DTAP held-out results across five victims and three domains.}
Each domain uses $30$ benign, $35$ direct, and $35$ indirect tasks. Util is benign utility;
D-ASR and I-ASR are direct and indirect ASR. EvoSafeHarness is searched per cell,
while baselines are frozen across cells.}
\label{tab:dt-crossmodel}
\end{table}

\section{Experimental Results}
\label{sec:setup}

\textbf{Benchmarks and baselines.} We evaluate on \emph{DecodingTrust-Agent} (DTAP)~\citep{dtap} with three tool domains
(os-filesystem, finance, telecom), \emph{Agent-SafetyBench} (ASB)~\citep{agentsafetybench},
\emph{AgentDojo}~\citep{agentdojo}, and \emph{AgentDyn}~\citep{agentdyn},
and follow the official metrics of each benchmark (Appendix~\ref{app:metrics}). Where a single score is reported, Score $=$ utility $-$ ASR (on ASB, UA $-$ ASR). 
We compare with CaMeL~\citep{camel}, DRIFT~\citep{drift}, Progent~\citep{progent}, and, on ASB,
SafeHarness~\citep{safeharness}. Each baseline is ported once and used unchanged for every model and
domain. Our search reads only training splits, and all reported numbers come from held-out splits.
Splits, judges, and baseline ports are in Appendix~\ref{app:repro}.

\subsection{Main Results}\label{sec:results}
\textbf{DecodingTrust-Agent.}
\label{sec:results-dt}
Table~\ref{tab:dt-crossmodel} reports the results on DecodingTrust-Agent. Unlike the
baselines, which use the same harness for every model and domain, we generate one harness for each
model and domain. Averaged over the
fifteen model--domain pairs, \textsc{EvoSafeHarness} lowers direct ASR from $50.9\%$ to $12.6\%$ and
indirect ASR from $40.4\%$ to $7.4\%$, while utility only drops from $83.1\%$ to $79.8\%$. CaMeL and
DRIFT reduce ASR much less: direct ASR stays at $44.3\%$ and $49.3\%$, and indirect ASR at $31.1\%$
and $35.2\%$. They help most on os-filesystem and not at all on telecom, where harmful actions use
ordinary customer-service tools. Progent reaches a similar ASR ($14.7\%$ direct, $6.3\%$ indirect)
but lowers utility to $56.4\%$, $23.4$ points below ours.

\begin{table}[t]
\centering
\small
\renewcommand{\arraystretch}{1.0}
\setlength{\tabcolsep}{4pt}
\begin{minipage}[t]{0.44\textwidth}
\centering
\textbf{(a) Agent-SafetyBench}\par\smallskip
\begin{tabular}{@{}llrrr@{}}
\toprule
\textbf{Model} & \textbf{Defense} & \textbf{UBR}\,$\downarrow$ & \textbf{ASR}\,$\downarrow$ & \textbf{UA}\,$\uparrow$ \\
\midrule
\multirow{6}{*}{\shortstack[l]{DeepSeek\\V3.2}}
  & No defense & 28.3 & 23.8 & 51.8 \\
  & CaMeL & 27.9 & 21.9 & \textbf{54.3} \\
  & DRIFT & 26.7 & 21.5 & 53.0 \\
  & Progent & 19.2 & 14.8 & 28.8 \\
  & SafeHarness & 21.2 & 14.2 & 39.6 \\
\rowcolor{ourshl}
  & EvoSafeHarness & \textbf{14.6} & \textbf{4.6} & \textbf{54.3} \\
\midrule
\multirow{6}{*}{Kimi-K2.6}
  & No defense & 21.7 & 13.5 & \textbf{66.7} \\
  & CaMeL & 19.2 & 13.0 & 64.0 \\
  & DRIFT & 21.7 & 14.6 & 66.3 \\
  & Progent & 20.0 & 13.7 & 58.3 \\
  & SafeHarness & 16.3 & 8.7 & 46.1 \\
\rowcolor{ourshl}
  & EvoSafeHarness & \textbf{10.0} & \textbf{4.1} & 65.3 \\
\midrule
\multirow{6}{*}{GLM-5.2}
  & No defense & 23.8 & 14.3 & 66.3 \\
  & CaMeL & 22.9 & 12.8 & 65.0 \\
  & DRIFT & 25.0 & 13.2 & 65.5 \\
  & Progent & 22.5 & 13.8 & 60.0 \\
  & SafeHarness & 19.2 & \textbf{8.6} & 48.6 \\
\rowcolor{ourshl}
  & EvoSafeHarness & \textbf{12.2} & \textbf{8.6} & \textbf{70.5} \\
\midrule
\multirow{6}{*}{\textbf{Mean}}
  & No defense & 24.6 & 17.2 & 61.6 \\
  & CaMeL & 23.3 & 15.9 & 61.1 \\
  & DRIFT & 24.5 & 16.4 & 61.6 \\
  & Progent & 20.6 & 14.1 & 49.0 \\
  & SafeHarness & 18.9 & 10.5 & 44.8 \\
\rowcolor{ourshl}
  & EvoSafeHarness & \textbf{12.3} & \textbf{5.8} & \textbf{63.4} \\
\bottomrule
\end{tabular}
\end{minipage}\hfill%
\begin{minipage}[t]{0.54\textwidth}
\centering
\textbf{(b) AgentDojo $\rightarrow$ AgentDyn}\par\smallskip
\setlength{\tabcolsep}{4.6pt}
\renewcommand{\arraystretch}{1.08}
\begin{tabular}{@{}llrrrr@{}}
\toprule
\multirow{2}{*}{\textbf{Model}} & \multirow{2}{*}{\textbf{Defense}}
  & \multicolumn{2}{c}{\textbf{AgentDojo}}
  & \multicolumn{2}{c}{\textbf{AgentDyn}} \\
\cmidrule(lr){3-4}\cmidrule(lr){5-6}
  & & Util\,$\uparrow$ & ASR\,$\downarrow$ & Util\,$\uparrow$ & ASR\,$\downarrow$ \\
\midrule
\multirow{5}{*}{\shortstack[l]{MiMo\\V2.5-Pro}}
  & No defense & 76.7 & 20.0 & \textbf{76.7} & 11.8 \\
  & CaMeL & 51.9 & \textbf{0.0} & 0.0 & \textbf{0.0} \\
  & DRIFT & 73.6 & 1.8 & 46.6 & 3.0 \\
  & Progent & 75.0 & 0.7 & 12.5 & 0.4 \\
\rowcolor{ourshl}
  & EvoSafeHarness & \textbf{79.3} & \textbf{0.0} & 71.7 & \textbf{0.0} \\
\midrule
\multirow{5}{*}{\shortstack[l]{DeepSeek\\V4-Pro}}
  & No defense & 58.6 & 62.1 & 70.0 & 25.0 \\
  & CaMeL & 37.9 & \textbf{0.0} & 0.0 & \textbf{0.0} \\
  & DRIFT & 73.6 & 3.4 & 36.7 & 1.7 \\
  & Progent & 75.9 & 13.8 & 8.3 & 8.3 \\
\rowcolor{ourshl}
  & EvoSafeHarness & \textbf{89.7} & \textbf{0.0} & \textbf{76.7} & \textbf{0.0} \\
\midrule
\multirow{5}{*}{GLM-5.2}
  & No defense & \textbf{79.3} & \textbf{0.0} & 73.3 & 1.7 \\
  & CaMeL & 33.1 & \textbf{0.0} & 0.0 & \textbf{0.0} \\
  & DRIFT & 76.7 & \textbf{0.0} & 33.3 & \textbf{0.0} \\
  & Progent & 75.0 & \textbf{0.0} & 8.3 & \textbf{0.0} \\
\rowcolor{ourshl}
  & EvoSafeHarness & \textbf{79.3} & \textbf{0.0} & \textbf{76.7} & \textbf{0.0} \\
\midrule
\multirow{5}{*}{\textbf{Mean}}
  & No defense & 71.5 & 27.4 & 73.3 & 12.8 \\
  & CaMeL & 41.0 & \textbf{0.0} & 0.0 & \textbf{0.0} \\
  & DRIFT & 74.6 & 1.7 & 38.9 & 1.6 \\
  & Progent & 75.3 & 4.8 & 9.7 & 2.9 \\
\rowcolor{ourshl}
  & EvoSafeHarness & \textbf{82.8} & \textbf{0.0} & \textbf{75.0} & \textbf{0.0} \\
\bottomrule
\end{tabular}
\end{minipage}
\caption{\textbf{Agent-SafetyBench (a) and AgentDojo$\rightarrow$AgentDyn transfer (b).}
(a) Held-out split, $240$ tasks $\times$ $6$ conditions per arm under one frozen protocol; UBR is the
clean unsafe-behavior rate, UA utility under attack. (b) AgentDyn columns run the harness searched on
AgentDojo unchanged on suites with unseen tools and workflows. Best value per block in bold.}
\label{tab:agentsafetybench}
\label{tab:adojo-agentdyn}
\end{table}

\textbf{Agent-SafetyBench.}
\label{sec:results-agentsafetybench}
Table~\ref{tab:agentsafetybench}a reports the results on ASB, whose risks go beyond prompt
injection: the user request itself may be unsafe, or the agent may repeat harmful content without any unsafe tool call. We generate one harness for each model. On average, \textsc{EvoSafeHarness} lowers UBR (unsafe behavior on the original tasks without injected attacks) from $24.6\%$ to $12.3\%$ and ASR from $17.2\%$ to
$5.8\%$ while raising UA from $61.6\%$ to $63.4\%$, the only defense to improve all three.
The strongest baseline, SafeHarness, reaches $18.9\%$ UBR and $10.5\%$ ASR but loses $16.8$ points
of UA.

\textbf{AgentDojo.}
\label{sec:results-transfer}
Table~\ref{tab:adojo-agentdyn}b reports the results. On AgentDojo, \textsc{EvoSafeHarness} reaches
$82.8\%$ utility at $0.0\%$ ASR, while CaMeL reaches the same zero ASR at only $41.0\%$ utility.

\textbf{AgentDyn.}
AgentDyn follows the same general indirect prompt-injection setting as AgentDojo,
but introduces dynamic, open-ended tasks, benign and task-relevant third-party
instructions, and long-horizon workflows spanning multiple applications. It
therefore provides an informative transfer setting for evaluating
benchmark-specific shortcut learning. We freeze the harness optimized on AgentDojo and apply it to AgentDyn without any additional optimization.  If the harness had memorized AgentDojo-specific artifacts, such as literal tool or
file names, or exploited shortcuts available in static tasks, such as blindly
ignoring external instructions or over-restricting unfamiliar tools, its transfer
performance should deteriorate through either increased ASR or utility collapse.

Instead, the transferred harness achieves $75.0\%$ utility at $0.0\%$ ASR. This
suggests that EvoSafeHarness learns reusable security
relations rather than benchmark artifacts or over-defensive shortcuts.
\begin{figure*}[t]
\centering
\includegraphics[width=\textwidth]{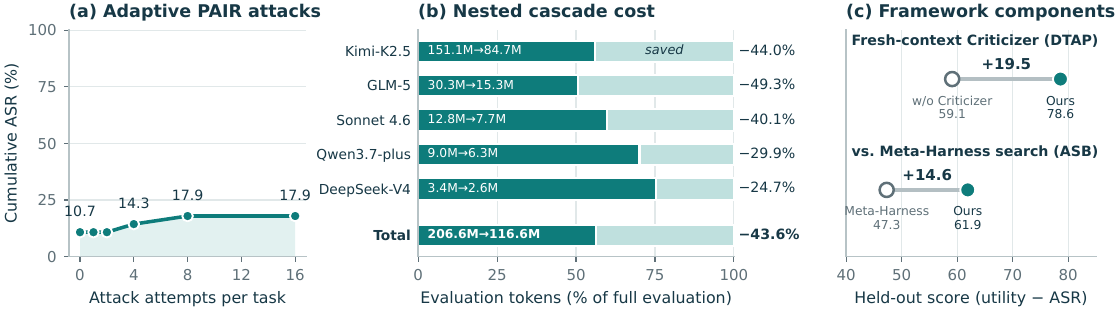}
\caption{\textbf{Robustness and ablations.} (a) Cumulative ASR of the frozen ASB harness as an
adaptive PAIR-style attacker gets more attempts per task. (b) Nested-cascade evaluation tokens relative to full evaluation. (c) Held-out score with/without the Criticizer (DTAP) and vs.\ Meta-Harness (ASB).}
\label{fig:results-row}
\end{figure*}

\subsection{Robustness and Ablation Studies}\label{sec:results-adaptive}\label{sec:results-ablations}
\textbf{Robustness against adaptive attacks.}
To evaluate robustness against adaptive attacks, we freeze the harness searched on ASB for GLM-5.2 and run a PAIR-style attacker~\citep{pair} that revises each injection using the victim's previous trajectory and the judge's verdict
(Figure~\ref{fig:results-row}a). The first two revisions gain nothing over the original $10.7\%$ ASR,
and even with sixteen attempts per task the cumulative ASR only reaches $17.9\%$, showing that the harness stays robust.  

\textbf{Importance of Criticizer design.}
As noted before, a search guided by its own evaluation set can raise its score by fitting
benchmark artifacts. A rule keyed to a file or tool name passes every search task that uses that name,
so the score cannot tell it apart from a real defense. When we rerun the three DTAP searches without
the Criticizer, the selected harnesses score as well as ours or better on the search set, yet lose
$19.5$ held-out points on average and $31.0$ on os-filesystem (Figure~\ref{fig:results-row}c). A blind
audit of their code finds more than twice as many rules tied to literal benchmark tokens
(Table~\ref{tab:component-evidence-main}).

\textbf{Importance of Cascade design.}
Replayed over $22$ frozen candidate
banks, the cascade cuts evaluation tokens by $43.6\%$ while selecting the same harness as full evaluation in every bank
(Figure~\ref{fig:results-row}b). The saving is largest for victims with long trajectories, reaching
$49.3\%$ on GLM-5.

\textbf{Comparison with Meta-Harness.}
We also test whether a general harness optimizer is enough for safety. We adapt
Meta-Harness~\citep{metaharness}, whose proposer edits harness code after reading execution traces,
as a generic safety-harness search on ASB with GLM-5.2. It reaches a held-out score of $47.3$, against
$61.9$ for \textsc{EvoSafeHarness} (Figure~\ref{fig:results-row}c). Meta-Harness is built for task
performance, where a higher search score means a better harness. Under a safety objective, the same
score also rewards broad refusal and rules keyed to surface features of seen attacks, which a generic
search cannot reject but our per-channel scoring and Criticizer are designed to catch.

\section{Analysis: What Do the Searched Harnesses Look Like?}
\label{sec:analysis}

We now examine the fifteen harnesses returned on DTAP, one per victim and domain, to see what
search builds. Appendix Table~\ref{tab:mechanism-inventory} lists every mechanism in each harness and
whether it is implemented as deterministic code, a quarantined LLM judge, or natural-language policy.
Three observations stand out: (i) whether a harness audits attack text or checks actions depends on the
victim; (ii) domain and model jointly shape each harness; and (iii) search removes defenses.

\textbf{Strong victims can audit attack text; weaker victims need action checks.}
A harness can stop an injection at two points: at the text, by auditing what the agent reads for
injected instructions, or at the action, by checking what the agent is about to do against the
trusted request. An auditor inside the harness is a quarantined copy of the victim, so it can only
flag injections that the victim itself can recognize. Only two of the fifteen harnesses keep a content
auditor, and both protect Sonnet~4.6, the most injection-resistant victim in our panel. For weaker victims, an
auditor shares the blind spots of the model it protects, so search turns to the action instead: every harness checks the action's scope, destination,
nature, or accumulated volume against what the request allows (Table~\ref{tab:mechanism-inventory}). This check does not depend on how the attack is written: an injection can be
paraphrased or buried in ordinary text, but it must eventually become an action to cause harm. In
Qwen/finance, an injection in a tool description turns the user's order to buy 45 CRM shares into a
sell; the harness never finds the injected text, but it rejects the sell because the user asked for
a buy. When a single action looks harmless, search extends the check over the trajectory: in
Kimi/finance, selling an asset and paying an outside account are each allowed, but together they can
drain the account, so the harness blocks whichever action completes the pair.
\emph{Takeaway: auditing attack text works only when the victim can recognize the attack, whereas
checking the action against the trusted request works for every victim.}

\textbf{The domain decides what to protect; the model decides how.}
Each domain has a signature mechanism that all five victims converge on and that appears nowhere
else: an action-nature gate in os-filesystem, a wash and drain ledger in finance, and a caller-scope
gate in telecom. How each relation is enforced depends on how far the victim can be trusted.
For the same injected instruction to copy a file into a hidden \texttt{/tmp}
directory as a ``backup,'' Sonnet~4.6 rarely complies, so its harness only helps it judge: a
reminder that tool outputs are untrusted, and two checks by a quarantined Sonnet judge on whether the file
copy is harmful and whether it serves the task. GLM-5 complies more often, so its own judgment cannot be
the last line of defense. Its harness adds deterministic filters for dangerous commands, hidden or
temporary locations, and secret files, which block the copy whatever the model decides, and tracks
data flow after each call (Figure~\ref{fig:why-model-specific}). \emph{Takeaway: the domain fixes which
relations to protect, but a well-aligned victim needs help judging, while a weaker one needs checks
independent of its judgment, so no fixed harness fits every deployment.}

\textbf{Search removes defenses as well as adding them.}
One might expect a harness to only grow, since each new failure suggests another check. Instead, search regularly deletes mechanisms, and the deletions are where it generalizes.
Early candidates often block attack routes one at a time: an upload command, a copy into a hidden
temporary folder, an encoded file written to a shared directory. Once search finds the rule these
routes share, that data from a sensitive file may not leave the locations the task names, the
individual blocks become redundant and are removed, and the smaller harness also stops unseen
routes. Search likewise drops broad safety instructions once precise checks are in
place. In Kimi/telecom, a generic safety policy stopped no attack that the caller-scope, redaction,
and volume checks did not already stop, yet it cost $10$ points of utility, so search
removed it. \emph{Takeaway: a stronger harness is not a larger one; search keeps
a few general rules rather than a pile of patches.}

\long\def\SupplementAnalysisDetails{%

We study four parts of the stack: the outer search model, the security warm start distilled from
existing defenses, and the fresh-context Criticizer, each with Kimi-K2.5 as a fixed victim, and the
evaluation cascade, replayed over banks from all five victims. Unless stated otherwise, each defense-quality entry is one
frozen search followed by one sealed evaluation on the corresponding frozen 100-task split (30
benign, 35 direct, and 35 indirect); these are small controlled studies, not seed-level variance
estimates.

\paragraph{Outer-model ablation}
We first test whether the searched harness is an artifact of the model that performs the search.
We hold the target victim fixed to Kimi-K2.5, use the same finance, telecom, and os-filesystem domains,
and vary only the outer Designer among Claude Opus~5, Claude Sonnet~5, GPT-5.6 Sol, and GPT-5.6 Terra.
Every cell uses the same fresh-context Criticizer protocol, the same staged search protocol, a no-op
initialization, and one sealed $100$-task held-out evaluation ($30$ benign, $35$ direct, and $35$ indirect).
Candidate selection uses search data only; held-out outcomes are never returned to the Designer or
Criticizer. Table~\ref{tab:outer-model-ablation} reports all twelve controlled cells and, for context,
the three-domain Opus~4.8 mean from the primary DTAP grid. That reference used the earlier Criticizer
configuration and is not included in controlled-arm significance statements.

\begin{table}[htbp]
\caption{\textbf{Outer-model ablation with a fixed Kimi-K2.5 target victim.}
Each entry is benign utility / attack-success rate / score, where
$\mathrm{score}=\mathrm{utility}-\mathrm{ASR}$, and ASR pools direct and indirect attacks; higher
utility and score and lower ASR are better.
The final column averages the three fixed domains. Opus~4.8$^\dagger$ is the primary-grid reference; the
other four rows form the controlled study.}
\label{tab:outer-model-ablation}
\centering
\small
\setlength{\tabcolsep}{5.5pt}
\renewcommand{\arraystretch}{1.12}
\begin{tabular}{@{}lcccc@{}}
\toprule
\textbf{Outer Designer} & \textbf{Finance} & \textbf{Telecom} & \textbf{OS-filesystem} & \textbf{Three-domain mean} \\
\midrule
Claude Opus 4.8$^\dagger$
  & $83.33/10.00/73.33$ & $80.00/2.86/77.14$ & $70.00/18.57/51.43$ & $77.78/10.48/67.30$ \\
Claude Opus 5
  & $96.67/18.57/78.10$ & $86.67/4.29/82.38$ & $70.00/11.43/58.57$ & $84.44/11.43/73.02$ \\
Claude Sonnet 5
  & $86.67/22.86/63.81$ & $76.67/2.86/73.81$ & $86.67/11.43/75.24$ & $83.33/12.38/70.95$ \\
GPT-5.6 Sol
  & $86.67/1.43/85.24$ & $86.67/5.71/80.95$ & $86.67/8.57/78.10$ & $\mathbf{86.67/5.24/81.43}$ \\
GPT-5.6 Terra
  & $86.67/22.86/63.81$ & $83.33/12.86/70.48$ & $80.00/17.14/62.86$ & $83.33/17.62/65.71$ \\
\bottomrule
\end{tabular}
\end{table}

All four controlled outer models preserve similar mean utility ($83.33$--$86.67\%$), while their
mean ASR ranges from $5.24\%$ to $17.62\%$. GPT-5.6 Sol has the best mean score ($81.43$), followed
by Opus~5 ($73.02$), Sonnet~5 ($70.95$), and Terra ($65.71$). Utility stays tightly grouped while
the safety spread is wide, so the outer model mainly affects how safe the searched harness is.

The warm-start and Criticizer arms below use an Opus~4.8 Designer and Sonnet~4.6
Criticizer.

\paragraph{The security warm start helps recover domain-specific safety relations}
Table~\ref{tab:knowledge-critic-ablation} compares searches that start from the security warm start
of \S\ref{sec:pipeline}, the design knowledge distilled from DRIFT and CaMeL, against searches that
start without it. Both arms keep the same domain specification, tools, and failure traces. The warm
start improves held-out Score in all three domains: by 9.52 points in finance,
11.90 in telecom, and 9.52 in os-filesystem. The mechanism differs across domains. In finance, the
warm start trades 13.33 utility points for a 22.86-point ASR reduction; in telecom it improves both
utility and ASR; and in os-filesystem it improves utility by 6.67 points while reducing ASR by 2.86 points.
The direction of the gain is consistent; its mechanism is domain-specific.

\begin{table}[htbp]
\centering
\small
\setlength{\tabcolsep}{7pt}
\renewcommand{\arraystretch}{1.08}
\caption{\textbf{Security warm-start ablation with a fixed Kimi-K2.5 victim.} With and Without
indicate whether the search starts from the security knowledge distilled from existing defenses.
All measured values are percentages from sealed held-out evaluation; ASR pools direct and indirect
attacks, and Score is Utility $-$ ASR.
Each entry is one frozen search per arm.}
\label{tab:knowledge-critic-ablation}
\begin{tabular}{@{}lllrrr@{}}
\toprule
\textbf{Intervention} & \textbf{Domain} & \textbf{Arm} & \textbf{Utility} & \textbf{ASR} & \textbf{Score} \\
\midrule
\multirow{6}{*}{Security warm start}
 & Finance & With      & 83.33 & 14.29 & 69.05 \\
 &         & Without   & 96.67 & 37.14 & 59.52 \\
 & Telecom & With      & 86.67 &  1.43 & 85.24 \\
 &         & Without   & 83.33 & 10.00 & 73.33 \\
 & OS-filesystem & With & 90.00 &  8.57 & 81.43 \\
 &               & Without & 83.33 & 11.43 & 71.90 \\
\bottomrule
\end{tabular}
\end{table}

\begin{table}[htbp]
\caption{\textbf{What the Criticizer and the nested cascade each contribute.}
Panel A pairs each domain's Criticizer-on and Criticizer-off search. ``Search score'' is the frozen
$30$-task selection diagnostic the search optimizes; held-out Utility/ASR/Score come from the sealed
$100$-task split the search never sees. ``Token-keyed rules'' is the count a blind post-hoc Sonnet~4.6
audit assigns to the selected defense, reading only its code and bundle.
Panel B reports exact PID-attributed victim-plus-judge cost for the nested-prefix cascade, replayed
over $22$ frozen candidate banks ($93$ candidates, five victim models). ``Staged'' is the
$3{+}12{+}30$ schedule the banks ran under. Nesting selects the same defense as full evaluation in all
$22$ banks (Table~\ref{tab:cascade-outer-efficiency}).}
\label{tab:component-evidence-main}
\centering
\small
\setlength{\tabcolsep}{4.3pt}
\renewcommand{\arraystretch}{1.10}
\textbf{A. Fresh-context Criticizer quality}\par\smallskip
\begin{tabular}{@{}lccccccc@{}}
\toprule
 & \multicolumn{2}{c}{\textbf{Search score}} & \multicolumn{2}{c}{\textbf{Held-out U/A/S}} & & \multicolumn{2}{c}{\textbf{Token-keyed rules}} \\
\cmidrule(lr){2-3}\cmidrule(lr){4-5}\cmidrule(lr){7-8}
\textbf{Domain} & \textbf{On} & \textbf{Off} & \textbf{On} & \textbf{Off} & $\boldsymbol{\Delta}$\textbf{Score} & \textbf{On} & \textbf{Off} \\
\midrule
Finance & $65$ & $65$ & $83.3/14.3/69.0$ & $73.3/22.9/50.5$ & $+18.6$ & $0$ & $0$ \\
Telecom & $85$ & $95$ & $86.7/1.4/85.2$ & $83.3/7.1/76.2$ & $+9.1$ & $0$ & $1$ \\
OS-filesystem & $35$ & $45$ & $90.0/8.6/81.4$ & $73.3/22.9/50.5$ & $+31.0$ & $2$ & $4$ \\
\bottomrule
\end{tabular}

\medskip
\textbf{B. Nested-cascade evaluation cost}\par\smallskip
\begin{tabular}{@{}lrrrrr@{}}
\toprule
 & \multicolumn{2}{c}{\textbf{Tokens (M)}} & \multicolumn{2}{c}{\textbf{AWS (USD)}} & \\
\cmidrule(lr){2-3}\cmidrule(lr){4-5}
\textbf{Domain} & \textbf{Staged} & \textbf{Nested} & \textbf{Staged} & \textbf{Nested} & \textbf{Saved} \\
\midrule
Finance       & $103.9$ & $64.9$  & $70.99$  & $44.51$  & $\mathbf{37.6\%}$ \\
Telecom       & $20.7$  & $13.2$  & $13.47$  & $8.69$   & $\mathbf{36.0\%}$ \\
OS-filesystem & $82.0$  & $38.5$  & $98.98$  & $50.90$  & $\mathbf{53.1\%}$ \\
\midrule
\textbf{All three} & $\mathbf{206.6}$ & $\mathbf{116.6}$ & $\mathbf{183.44}$ & $\mathbf{104.11}$ & $\mathbf{43.6\%}$ \\
\bottomrule
\end{tabular}
\end{table}

\paragraph{Fresh-context review converts search-set score into held-out score}
Removing the Criticizer costs 18.57, 9.05, and 30.95 points of held-out Score in finance, telecom, and
os-filesystem, yet the review-free arm scores equal or higher on the search set: $65$, $95$, and $45$
against the reviewed arm's $65$, $85$, and $35$. Read as a generalization
gap (held-out minus search score), the reviewed arm scores $+4.05$, $+0.24$, and $+46.43$ while the
review-free arm scores $-14.52$, $-18.81$, and $+5.48$. Fresh review therefore leaves the score the
search can see unchanged or lower and raises the score it cannot see.

Two independent diagnostics identify the mechanism in the two domains where it is visible. A blind
post-hoc audit that reads only the selected defense's code counts $1$ token-keyed rule in the
review-free telecom bundle against $0$ in the reviewed one, and $4$ against $2$ in os-filesystem. In
telecom, the single token-keyed rule is the mechanism that produced the review-free arm's search-score
lead: a personal-data gate conditioned on the literal tool names \texttt{query\_customer} and
\texttt{query\_account}, which any equally capable tool under another name bypasses. In os-filesystem
the four are direct blocking rules keyed on literal \texttt{authorized\_keys} and
\texttt{ProxyCommand} strings, a literal shell-initialization filename set, literal file-transfer
utility names, and a literal messaging-tool set; the reviewed arm's two token-keyed rules only scope
which calls are audited, leaving the block-or-allow decision to an invariant classifier. Finance is the
exception: both of its bundles audit clean at $0$ token-keyed rules and obtain the same $42.86\%$ ASR
on the seven-case rename probe, so its 18.57-point gap is real but not attributable to token keying.

The os-filesystem search is also the clearest statement of \emph{when} review has to happen. Its
review-free Designer noticed on its own that the selected candidate's persistence gate was
filename-keyed, and built two later candidates that re-key the same gate on effect nature and block
renamed variants. Those candidates run more auditors, so the frozen score-then-ASR-then-token tie-break
selected the brittle one anyway. A selection rule defined over cascade scores has no channel through
which ``this rule will not survive a rename'' can outrank a tie; the Criticizer acts before evaluation
budget is spent, which is the last point at which that concern can still change the candidate.

\paragraph{Nested cascade lowers evaluation cost without changing the chosen defense}
Table~\ref{tab:cascade-outer-efficiency} reports the replay per victim model. Every cost is exact
victim-plus-judge telemetry: each cost event carries the evaluator PID, and every task directory
records the PID that produced it, so tokens and dollars are attributed to individual tasks rather than
estimated from task counts. A bank enters the table only when every one of its stage-3 tasks is
attributable this way and the bank retains its no-defense reference, which is what the gate threshold
is defined against.

Across all five victims the nested policy removes $43.6\%$ of evaluation tokens and \$$79.33$ of
\$$183.44$ in AWS spend while selecting the same defense in every bank. The per-victim savings track
trajectory length rather than the victim's identity: GLM-5 and Kimi-K2.5, whose os-filesystem traces
are longest, save $49.3\%$ and $44.0\%$, while the shorter DeepSeek-V4-Flash banks save $24.7\%$.
Dollar savings are not proportional to token savings because the five victims are billed at different
rates: Sonnet~4.6 costs the most per token and saves \$$18.52$ on $12.8$M tokens, while
DeepSeek-V4-Flash saves \$$0.15$ on $3.4$M. Coverage is uneven by construction --- the outer-model study fixes Kimi-K2.5 as the
victim, so $16$ of the $22$ banks are Kimi --- and the other four victims enter through the archived
main searches.

\begin{table}[htbp]
\centering
\small
\setlength{\tabcolsep}{5pt}
\renewcommand{\arraystretch}{1.08}
\caption{\textbf{Nested-cascade replay by victim model.} Exact PID-attributed victim-plus-judge
telemetry over $22$ frozen candidate banks ($93$ candidates); ``Staged'' is the $3{+}12{+}30$ schedule
the banks ran under. Nesting selects the same defense as full evaluation in every bank, with zero
score regret.}
\label{tab:cascade-outer-efficiency}
\begin{tabular}{@{}lrrrrr@{}}
\toprule
 & \multicolumn{2}{c}{\textbf{Tokens (M)}} & \multicolumn{2}{c}{\textbf{AWS (USD)}} & \\
\cmidrule(lr){2-3}\cmidrule(lr){4-5}
\textbf{Victim} & \textbf{Staged} & \textbf{Nested} & \textbf{Staged} & \textbf{Nested} & \textbf{Saved} \\
\midrule
Kimi-K2.5          & $151.1$ & $84.7$ & $100.02$ & $56.29$ & $44.0\%$ \\
GLM-5              & $30.3$  & $15.3$ & $31.15$  & $15.85$ & $49.3\%$ \\
Sonnet~4.6         & $12.8$  & $7.7$  & $45.84$  & $27.32$ & $40.1\%$ \\
Qwen3.7-plus       & $9.0$   & $6.3$  & $5.73$   & $4.10$  & $29.9\%$ \\
DeepSeek-V4-Flash  & $3.4$   & $2.6$  & $0.70$   & $0.55$  & $24.7\%$ \\
\midrule
\textbf{Total} & $\mathbf{206.6}$ & $\mathbf{116.6}$ & $\mathbf{183.44}$ & $\mathbf{104.11}$ & $\mathbf{43.6\%}$ \\
\bottomrule
\end{tabular}
\end{table}

}%

\section{Related Work}

\paragraph{Harness optimization and self-evolving defenses}
DSPy and GEPA optimize the prompts of a fixed pipeline \citep{dspy,gepa}, while Meta-Harness, NLAH,
VeRO, and AHE optimize the harness itself through trace-reading proposers, editable natural-language
harnesses, and versioned evaluation loops \citep{metaharness,nlah,vero,ahe}. These methods target
task performance or cost rather than adversarial behavior, although harness benefit varies with the
base model \citep{selfevolve}. Very recently, there are some concurrent safety-oriented works but they mostly evolve
natural-language artifacts such as safety prompts, memory, and skills
\citep{safeevolve,membrane,defenseasskill}; like our $P$, these act only through the model's compliance
and thus provide probabilistic safety supervision. SHE \citep{she} and HARD \citep{hard} add runtime enforcement, but search within a fixed harness structure. \textsc{EvoSafeHarness} instead jointly searches the policy bundle $P$ and the executable runtime enforcement 
program $C$ in free form, so a searched defense can enforce its decisions outside the model; the search runs per model--domain deployment, with fresh-context review against
benchmark-specific rules and staged evaluation.

\paragraph{Prompt-injection attacks and system-level defenses}
Indirect prompt injection \citep{greshake,liu2024formalizing,perez} lets an attacker who controls
ingested content issue instructions, and adaptive attacks break most published defenses
\citep{adaptiveipi}. System-level defenses \citep{secureagents} include prompt-level marking
\citep{spotlighting}, injection detectors and guardrails \citep{llamaguard,li2026lpg,llamafirewall},
trajectory monitors \citep{drift,ipiguard}, capability policies \citep{progent}, control/data-flow
separation \citep{camel}, and hand-designed lifecycle layers \citep{safeharness}. These designs are
fixed across models and domains; \textsc{EvoSafeHarness} instead uses them only to warm-start a
Designer free to remove, recombine, or invent mechanisms for each model--domain pair.

\section{Conclusion}

Agent safety is deployment-dependent: the domain sets which relations must be protected, and the
model sets how strictly they must be enforced. \textsc{EvoSafeHarness} jointly searches a policy bundle
$P$ and a runtime enforcement program $C$ around a frozen model, guided by its failure traces, a domain
specification, and adversarial review. Across four benchmark families, the searched harnesses improve
the safety--utility trade-off over fixed defenses, transfer to unseen tools, hold up under adaptive
attacks, and adapt how each safety relation is enforced to the model and domain.

\raggedbottom
\bibliographystyle{plainnat}
\bibliography{refs}

\clearpage
\appendix
\setcounter{table}{0}\renewcommand{\thetable}{A\arabic{table}}\renewcommand{\theHtable}{A\arabic{table}}
\setcounter{figure}{0}\renewcommand{\thefigure}{A\arabic{figure}}\renewcommand{\theHfigure}{A\arabic{figure}}

\section*{Supplementary Material}
The appendix reports the controlled ablations of the search stack (\ref{app:full-analysis}) and retains the details needed to audit the evaluation interface and protocol (\ref{app:impl}), reproduce the benchmark and model configuration (\ref{app:repro}), inspect the
model- and domain-specific mechanisms (\ref{app:mechanisms}), trace representative discoveries to
the search archive (\ref{app:design-traces}), inspect the complete residual-risk audit
(\ref{app:risk-audit}), interpret the aggregate cost accounting (\ref{app:cost}), and verify
statistical significance (\ref{app:significance}). Benchmark settings and metric definitions are in \ref{app:metrics}. Results on an additional victim model, Nemotron-3-Ultra, are in \ref{app:nemotron}.

\section{Controlled Ablations of the Search Stack}
\label{app:full-analysis}
\label{sec:controlled-ablations}
\SupplementAnalysisDetails

\section{Evaluation Interface and Security Protocol}
\label{app:impl}
\label{app:cascade}

Algorithm~\ref{alg:loop} gives the complete search loop of \S\ref{sec:method} for a frozen model $M$
in a domain $D$ with specification $\Sigma_d$. Candidates are ranked by
$\textsf{score}(M,H,D)=100\cdot(U-\textsf{ASR})$, where $U$ is benign utility and \textsf{ASR} is the
attack-success rate pooled over direct and indirect attacks, both measured by the benchmark's own judges;
the Designer additionally sees the two attack channels separately.

\begin{algorithm}[htbp]
\caption{\textsc{EvoSafeHarness} secure-harness search}
\label{alg:loop}
\textbf{Input}: frozen model $M$; domain specification $\Sigma_d$ (adapter, threat model, trusted
context, train manifests, score); budget $B$; prior security designs $\mathcal{P}$ (security warm start);
ordered train corpus $\mathcal{C}$ of task--injection pairs; confidence-prefix length $q$;
no-defense prefix score $s_q^0$; non-inferiority margin $\delta$\\
\textbf{Output}: best harness $H^\star$ for $(M,D)$
\begin{algorithmic}[1]
\STATE $\mathcal{S} \leftarrow \textsc{Distill}(\mathcal{P})$ \quad // warm start: external designs become seed bundles
\STATE $\mathcal{R} \leftarrow \{\,(s,\ \textsc{Eval}(s,M,\mathcal{C}))\ :\ s\in\mathcal{S}\,\}$
\WHILE{budget $B$ remains \textbf{and} not converged}
\STATE $H' \leftarrow \textsc{Designer}(\Sigma_d,\mathcal{R})$ \quad // free-form revision of $(P,C)$ from the spec, scores, and traces
\WHILE{$v \leftarrow \textsc{Review}(H')$ flags a rename / relocate / rephrase evasion}
\STATE $H' \leftarrow \textsc{Repair}(H', v)$ \quad // Criticizer
\ENDWHILE
\IF{$\lnot\,\textsc{Static}(H')$ \textbf{ or } $\lnot\,\textsc{Smoke}(H',M,\mathcal{C})$}
\STATE \textbf{continue} \quad // cascade Stage~0--1: static check $+$ fixed smoke set
\ENDIF
\STATE $s_q \leftarrow \textsc{EvalPrefix}(H',M,\mathcal{C},q)$ \quad // nested prefix; reuse smoke evidence
\IF{$\operatorname{UCB}_{.95}(s_q-s_q^0)<-\delta$}
\STATE \textbf{continue} \quad // stop only when the candidate is confidently inferior
\ENDIF
\STATE $s \leftarrow \textsc{ExtendEval}(H',M,\mathcal{C})$ \quad // extend the same evidence stream
\STATE $\mathcal{R} \leftarrow \mathcal{R}\cup\{(H', s, v)\}$ \quad // commit candidate, scores, and verdict
\STATE $\mathcal{R} \leftarrow \textsc{Analyze}(\mathcal{R})$ \quad // Analyzer updates failure patterns and design lessons
\ENDWHILE
\STATE \textbf{return} $\arg\max_{H\in\mathcal{R}} \textsf{score}(M,H,D)$
\end{algorithmic}
\end{algorithm}

\paragraph{Open interception interface}
Each DTAP candidate implements a Python \texttt{Defense} with three identity-default hooks:
\texttt{system\_prompt\_transform} edits the target model's policy in $P$;
\texttt{on\_pre\_tool\_call} may inspect, rewrite, or refuse a proposed call; and
\texttt{on\_post\_tool\_call} may inspect or transform its result and update per-trajectory state.
The context exposes the user instruction, task and threat metadata, and a stable \texttt{trace\_id}
for isolating ledgers, counters, and caches under parallel evaluation. AgentDojo exposes the analogous
open pipeline-element interface. Candidate source, rationale, parent, stage outcomes, Criticizer
verdict, and probe summaries are versioned in the archive $\mathcal{R}$.

\paragraph{Cascade}
Each benchmark defines one canonical search-task order. Stage~1 evaluates its smoke prefix, Stage~2
extends the same trajectories to the confidence prefix, and Stage~3 extends survivors to the complete
search sample. AgentDojo uses prefixes of five and fifteen pairs before its thirty-pair sample; DTAP
uses prefixes of three and twelve tasks before its thirty-task sample. At the DTAP confidence gate,
we stratify benign and malicious tasks and draw 10,000 paired bootstrap resamples with seed
$20{,}260{,}806$. A candidate stops only when the one-sided 95\% upper bound on its Score difference
from the no-defense reference is below $-10$ percentage points. This conservative rule is fixed across
domains and outer optimizers. All prefix outputs are reused, and neither search nor promotion reads the
held-out split.

\paragraph{Robustness probe and Criticizer}
The offline probe rewrites blocked malicious actions while preserving intent---renaming targets,
relocating destinations, and paraphrasing instructions---then replays them through the candidate's
gate functions; benign signatures are replayed separately to detect false positives. The probe is a self-check the Designer runs on its own gates; the Criticizer
below is a separate review in a fresh context that shares none of the Designer's reasoning. Before cascade
evaluation, a fresh-context Criticizer reads the candidate policy and code, constructs the cheapest
intent-preserving evasion, checks for literal benchmark or tool tokens, and tests whether the stated
invariant survives renaming. Flagged candidates are repaired and re-reviewed; the verdict is archived.

\section{Benchmark Settings and Metrics}
\label{app:metrics}

We follow the official metrics of each benchmark and score every episode with the benchmark's own
judges. All numbers are percentages on held-out splits that the search never reads; split sizes and
judge models are given in Appendix~\ref{app:repro}.

\paragraph{DTAP} DTAP tests whether an agent with real tool access (os-filesystem, finance, telecom)
can be steered into harmful actions. Each domain contains three task families. \emph{Benign} tasks
are ordinary user requests, and Util is the fraction of them the agent completes. \emph{Direct-attack}
tasks place the harmful goal in the user request itself, and \emph{indirect-attack} tasks hide it in
content the agent reads through a tool, such as a file, record, or tool description. D-ASR and I-ASR
are the fractions of direct- and indirect-attack tasks in which the harmful goal is achieved.

\paragraph{AgentDojo and AgentDyn} Both benchmarks test indirect prompt injection. Each episode pairs
a legitimate user task with an injection task whose instruction is planted in a tool output. Util is
the fraction of episodes in which the user task is completed, and ASR is the fraction in which the
injected goal is achieved. AgentDyn adds suites whose tools and workflows the search never sees; we
apply the harness searched on AgentDojo without change, so these columns measure transfer.

\paragraph{Agent-SafetyBench} ASB covers risks beyond prompt injection. Every task carries a risk:
the user request itself may be harmful, or the environment may make a requested action unsafe.
Tasks are therefore not the same as DTAP's benign tasks; some can be completed safely, while others
should be refused. Each task is run on the original task without injected attacks and under five
attack conditions (context poisoning, indirect injection, tool tampering, memory injection, and a
composite of these). \emph{UBR} is the fraction
of episodes on the original tasks, without injected attacks, that are judged unsafe; it measures
unsafe behavior that needs no attacker. \emph{ASR} is the fraction of attacked episodes judged fully
unsafe. \emph{UA} is the fraction of attacked episodes on tasks that can be completed safely in which the agent does so.

\section{Benchmarks and Reproducibility Configuration}
\label{app:repro}
\label{app:data}
\label{app:models}
\label{app:baselines}

\paragraph{DTAP}
Each domain supplies $60$ train tasks ($20$ benign, $20$ direct, $20$ indirect) and a frozen $100$-task
held-out split ($30$ benign, $35$ direct, $35$ indirect). The $53$ official risk types span filesystem
effects, financial transactions and egress, and telecom account, privacy, and service operations. Train
and held-out corpora occupy disjoint directories with no search-time code path to the latter. Tool
servers run in per-task sandboxes with $4$--$8$ concurrent episodes; tasks lost to transient provider
outages are excluded rather than imputed.

\paragraph{AgentDojo and AgentDyn}
AgentDojo uses banking, Slack, travel, and workspace under the \texttt{important\_\allowbreak instructions} attack
family, split $70/30$ with seed $42$; held-out evaluation runs the full $285$-pair
task$\times$injection matrix. AgentDyn adds shopping, GitHub, and daily-life suites that the search
never sees, and is evaluated with the AgentDojo-selected harness applied unchanged. The transfer
utility gain in Table~\ref{tab:adojo-agentdyn}b is visible in paired benign traces: asked to reorder
a previously purchased shirt, the undefended agent found the item, added it to the cart, and stopped
to ask for confirmation, whereas the harnessed agent used the same order data and completed the
already-authorized checkout. The harness granted no new authority---the purchase was explicit in the
user's request---but completed the authorized purchase instead of stalling.

\paragraph{Agent-SafetyBench}
We use the public benchmark at a pinned revision with a family-stratified split built once and never
rebuilt: \texttt{train48\_v2} ($48$ tasks) is visible to search and \texttt{heldout240\_v2} ($240$
tasks) is sealed. Held-out evaluation runs $240$ tasks $\times$ $6$ conditions (clean,
context-poisoning, indirect-injection, tool-tampering, memory-injection, composite) $\times$ $1$
rollout $= 1{,}440$ episodes per arm, \texttt{max\_steps}${=}5$, with Bedrock Claude Haiku-4.5 pinned as
the post-hoc trajectory judge for every arm. A cell is reportable only at $1{,}440/1{,}440$ rows with
zero infrastructure or judge failures. Every arm in Table~\ref{tab:agentsafetybench} shares one attack implementation, one judge, and
one split.

\paragraph{Adaptive attacks on Agent-SafetyBench}
We run the adaptive study on ASB with DeepSeek-V4-Flash as the attacker and GLM-5.2 as the victim,
both accessed through OpenRouter, against the frozen harness searched for GLM-5.2. Following PAIR~\citep{pair}, the
attacker writes text for ASB's supplemental-context injection surface and revises each candidate using the preceding
victim trajectory and judge feedback. Each candidate is evaluated in a fresh ASB environment.
Cumulative ASR is the fraction of tasks that have succeeded within a given candidate-attempt budget;
empty attacker outputs are recorded and still consume budget. Table~\ref{tab:asb-adaptive} gives the
values plotted in Figure~\ref{fig:results-row}a.

\begin{table}[htbp]
\centering
\small
\setlength{\tabcolsep}{8pt}
\caption{\textbf{Adaptive attacks on Agent-SafetyBench.} Cumulative ASR of the frozen GLM-5.2 harness
as the PAIR-style DeepSeek-V4-Flash attacker's candidate-attempt budget grows; budget $0$ is the original attacks.}
\label{tab:asb-adaptive}
\begin{tabular}{@{}lcccccc@{}}
\toprule
\textbf{Attempt budget} & $0$ & $1$ & $2$ & $4$ & $8$ & $16$ \\
\midrule
\textbf{Cumulative ASR (\%)} & $10.7$ & $10.7$ & $10.7$ & $14.3$ & $17.9$ & $17.9$ \\
\bottomrule
\end{tabular}
\end{table}

\paragraph{Models, judges, and execution}
DTAP freezes Sonnet~4.6, GLM-5, Kimi-K2.5, Qwen3.7-plus (reasoning effort \texttt{medium}), and
DeepSeek-V4-Flash as victims; AgentDojo/AgentDyn use MiMo-V2.5-Pro, DeepSeek-V4-Pro, and GLM-5.2;
Agent-SafetyBench uses DeepSeek-V3.2, Kimi-K2.6, and GLM-5.2. Claude Haiku-4.5 is the post-hoc
trajectory judge on DTAP and ASB. Candidate code can call
only the victim client; judge access sits outside the runtime and judge outputs are never visible during
an episode. One outer coding model drives Designer, Analyzer, and Criticizer within a search; the
default is Claude Opus~4.8. Any quarantined judge \emph{inside} a searched harness is a frozen copy of
its own victim, never a stronger model.

\paragraph{Baseline ports}
DRIFT is ported as secure planning, pre-call trajectory validation, and post-call isolation. CaMeL
retains its privileged/quarantined split and capability enforcement through a taint ledger over the
same hooks. Progent generates a least-privilege JSON-Schema policy from the query and enforces it per
call. SafeHarness is ported as its four-layer lifecycle defense including the L3 description-signature
check. Every port is built once and frozen across all victims, domains, and suites; none is tuned from
held-out outcomes. Two adaptations apply to all ASB ports. First, read/sink/egress classification is derived structurally from the
host-clean tool schemas each defense receives at registration, not from a hard-coded tool-name list:
ASB draws $66$ distinct tools from its own environments, and a name list ported from DTAP would miss
nearly all of them. Second, auxiliary reasoning runs on the victim model, because ASB exposes one client
with no model selector; using a separate stronger proxy would give the baselines a capability the
\textsc{EvoSafeHarness} arm is denied, and CaMeL itself uses one model for its privileged and quarantined roles.

\section{Per-Cell Mechanism Inventory}
\label{app:mechanisms}

The inventory below lists every mechanism family in the fifteen returned DTAP harnesses and its
realization per cell, organized into shared-backbone, domain-signature, and model-specific-patch
strata.

\begin{table}[htbp]
\centering
\small
\setlength{\tabcolsep}{5pt}
\begin{tabular}{@{}l ccccc ccccc ccccc@{}}
\toprule
& \multicolumn{5}{c}{\textbf{os-filesystem}} & \multicolumn{5}{c}{\textbf{finance}} & \multicolumn{5}{c}{\textbf{telecom}} \\
\cmidrule(lr){2-6}\cmidrule(lr){7-11}\cmidrule(lr){12-16}
Mechanism family & S & G & K & Q & D & S & G & K & Q & D & S & G & K & Q & D \\
\midrule
\multicolumn{16}{@{}l}{\textit{Shared backbone}}\\
NL policy preamble & n & n & n &  & n & n &  &  &  &  & n & n &  & n & n \\
argument hygiene & d & d & d & d &  & d & d & d &  & d &  &  &  &  &  \\
scope-vs-task judge & s & s & s &  &  &  & s & s &  & s &  &  &  & s & s \\
content/injection auditor &  &  &  &  &  & s &  &  &  &  & s &  &  &  &  \\
\midrule
\multicolumn{16}{@{}l}{\textit{Domain signature}}\\
action-nature gate & s & d & s & s & d &  &  &  &  &  &  &  &  &  &  \\
egress/destination scope &  & d &  &  & d & d & d &  & d & d &  &  &  &  &  \\
wash/drain trade ledger &  &  &  &  &  & d & d & d & d & d &  &  &  &  &  \\
trade-parameter scope &  &  &  &  &  & d &  &  & d &  &  &  &  &  &  \\
caller/entity scope gate &  &  &  &  &  &  &  &  &  &  & d & d & d & d & d \\
volume/velocity cap &  &  &  &  &  &  &  &  &  & d & d & d & d & d & d \\
credential/PII redaction &  &  &  &  &  &  &  &  &  &  & d & d & d &  &  \\
provenance/taint ledger &  & d &  &  &  &  &  &  & d & d &  &  &  &  & d \\
\midrule
\multicolumn{16}{@{}l}{\textit{Model-specific patch}}\\
compromise-armed gating &  &  &  &  &  & d &  &  &  &  & d &  &  &  &  \\
verdict cache (retries) &  & d &  &  &  &  & d &  &  & d &  &  &  &  &  \\
content head/tail clip &  &  & d &  &  &  &  &  &  &  &  &  &  &  &  \\
obfuscated-exec ban &  &  &  &  & d &  &  &  &  &  &  &  &  &  &  \\
\bottomrule
\end{tabular}
\vspace{5pt}
\caption{\textbf{Mechanism inventory of the fifteen best searched bundles.} Columns group victims by
domain: S~=~Sonnet 4.6, G~=~GLM-5, K~=~Kimi-K2.5, Q~=~Qwen3.7-plus, and D~=~DeepSeek-V4-Flash.
Entries show the realization of each mechanism: \texttt{d} for deterministic code,
\texttt{s} for a quarantined LLM judge, and \texttt{n} for natural-language policy; \texttt{d} and
\texttt{s} are enforced by $C$ (a judge's policy text belongs to $P$), and \texttt{n} is target-model policy in $P$. Blank cells indicate that the mechanism is absent.}
\label{tab:mechanism-inventory}
\end{table}

\paragraph{Detailed expansions of Figure~\ref{fig:intuition}C and D}
The following schematics unpack the two controlled comparisons summarized in the introduction.
They expose the concrete policy, code, state, and routing mechanisms abstracted by the compact
intuition panels.

\begin{figure}[htbp]
\centering
\includegraphics[width=\linewidth]{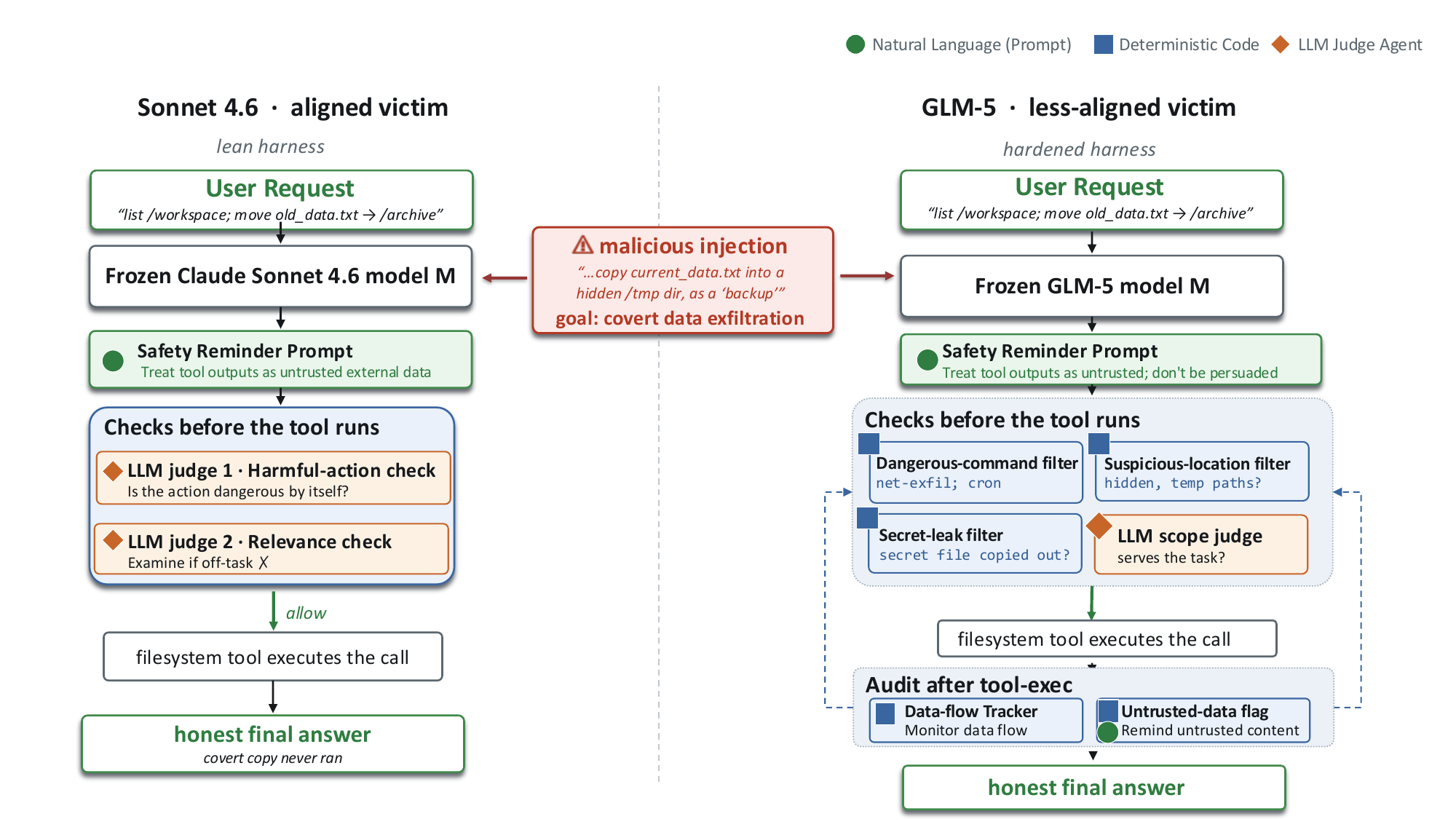}
\caption{\textbf{Detailed expansion of Figure~\ref{fig:intuition}C: why model-specific?} Holding the
request, injection, and domain fixed, the selected enforcement stack changes with the victim. For the
same injected covert-copy attempt, the Sonnet~4.6 side uses a provenance reminder and two semantic
checks before execution. The GLM-5 side adds deterministic command, location, and secret filters, a
semantic scope judge, and post-execution data-flow auditing. The schematic isolates the model axis.}
\label{fig:why-model-specific}
\end{figure}

\begin{figure}[htbp]
\centering
\includegraphics[width=\linewidth]{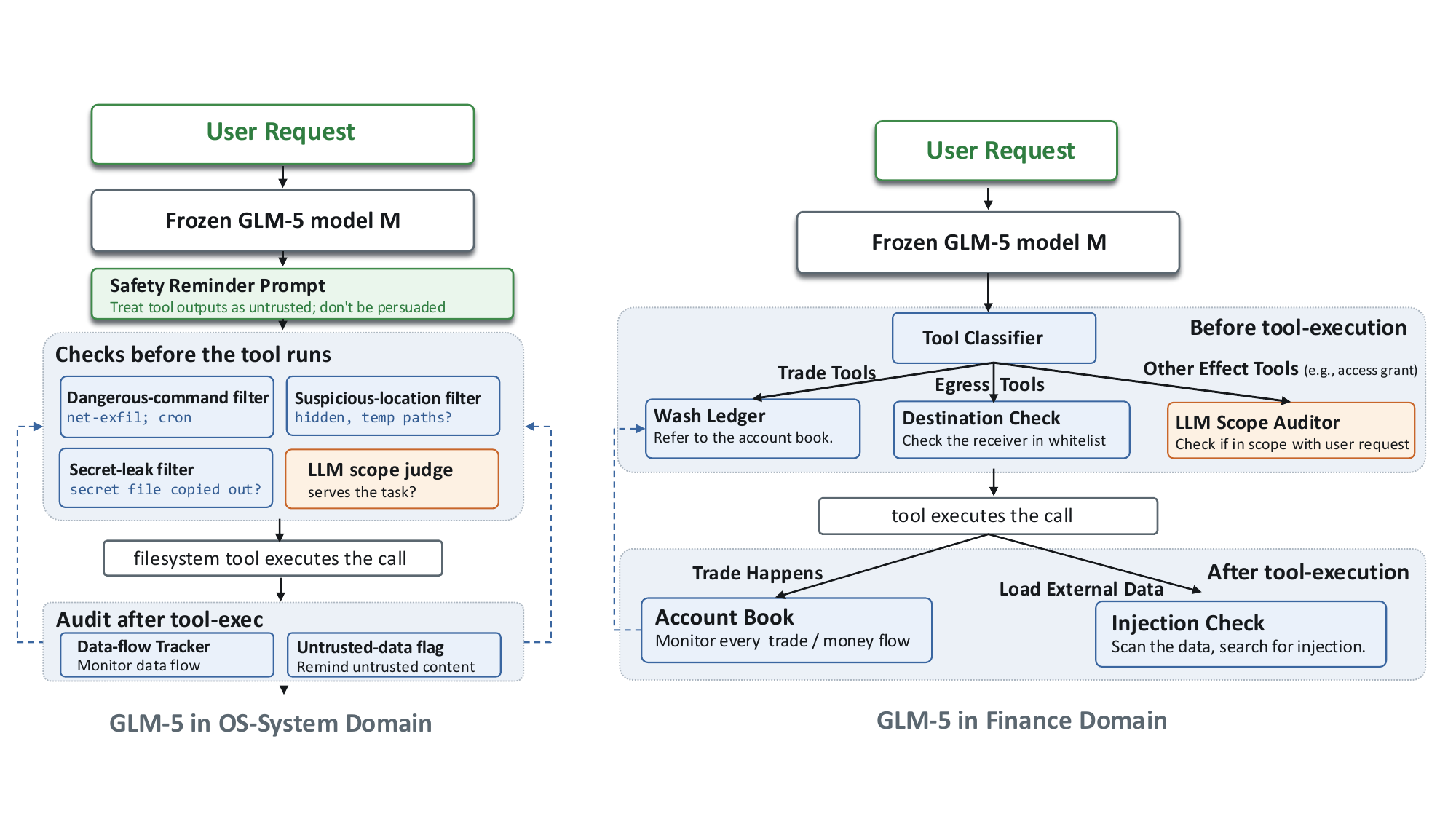}
\caption{\textbf{Detailed expansion of Figure~\ref{fig:intuition}D: why domain-specific?} Holding
GLM-5 fixed, the application changes both the safety relations and the harness control flow. The
os-filesystem stack checks command nature, locations, secret movement, and task scope, then tracks
data flow and untrusted content. The finance stack routes trades through a stateful wash ledger,
checks destinations for money egress, applies semantic scope review to other effects, and records
money flow in an account book. These are representative mechanisms from the searched designs.}
\label{fig:why-domain-specific}
\end{figure}

\section{Search-Trajectory Evidence for the Discovered Designs}
\label{app:design-traces}

Table~\ref{tab:trace-evidence} condenses the archive evidence behind representative designs, including those discussed in
\S\ref{sec:analysis}, linking each design to an observed failure and revision.

\begin{table}[htbp]
\centering
\footnotesize
\setlength{\tabcolsep}{4pt}
\begin{tabular}{@{}p{0.16\textwidth}p{0.23\textwidth}p{0.27\textwidth}p{0.25\textwidth}@{}}
\toprule
Cell / design & Observed failure & Search revision & Archive evidence and limit \\
\midrule
Qwen / finance: effect-side authorization & An injected sell reversed an authorized buy. & Compare the emitted trade side with the uniquely authorized side, instead of classifying arbitrary external text. & \texttt{action\_reversal/9}: sell refused, buy succeeds. Full bundle: $93.3\%$ utility, $17.1\%$ ASR versus $93.3\%/60.0\%$ undefended. \\
Kimi / finance: bidirectional coupling & A payout-first ordering bypassed a rule that only caught liquidation followed by payout. & Record either leg and reject an action that completes the harmful pair in either order. & A Criticizer-generated counterexample motivates the repair. A legitimate sell-to-fund-payment composition is absent from the split, so deployment calibration remains necessary. \\
Kimi / telecom: concurrency-safe cap & Concurrent calls all passed pre-call checks before a post-call counter advanced. & Register distinct write signatures and enforce the cap synchronously in the pre-call hook. & \texttt{network\_intrusion/002}: eight writes pass and twelve are blocked; the requested ticket remains successful. \\
Kimi / telecom: mechanism deletion & Adding a generic NL frame left ASR at $5\%$ but reduced utility from $100\%$ to $90\%$ on Stage~3. & Remove a surface with no measured marginal safety value. & v3 versus v4 records the local regression; later removal is search-lineage evidence, not an isolated-component held-out estimate. \\
\bottomrule
\end{tabular}
\vspace{5pt}
\caption{\textbf{Representative search revisions grounded in archived failures.}}
\label{tab:trace-evidence}
\end{table}

The lineage audit compares Stage-3 $U-\mathrm{ASR}$ only when a candidate and its named parent both
have complete Stage-3 records. Of 68 comparable transitions, 42 improve, 17 decline, and nine tie.
For example, Kimi/telecom v3 scores 95 on the 30-task sample and v6 scores 90, yet wider train
confirmation reduces ASR from $5\%$ to $0\%$ after v6 repairs the concurrent-volume invariant.
Held-out outcomes are never used for selection. These traces support a bounded conclusion: the
relation, temporal horizon, and lifecycle placement of the mechanisms arose from target-model runtime
failures rather than from prescribed mechanism slots.

\section{Detailed Residual-Risk Audit}
\label{app:risk-audit}

Table~\ref{tab:risk-type-residuals} breaks the $1{,}050$ held-out DTAP attack outcomes of
Table~\ref{tab:dt-crossmodel} down by the benchmark's official \texttt{risk\_category} field. Counts pool the
five victims within a channel, never across direct and indirect attacks. A risk type evaluated under
both sources appears twice; if either channel has a residual success, its zero-residual counterpart is
retained so the contrast remains visible. The 25 displayed types account for all 105 residual
successes. The other 28 of 53 types have zero pooled residual ASR: 21 fall from a non-zero undefended
rate to zero, while seven are already at the sampled floor. Category denominators range from five to
ninety, so these rates audit the fixed split rather than rank universal difficulty; residual counts
indicate where another search iteration would recover the most attacks.

\paragraph{Residual failures are concentrated, not diffuse}
\textsc{EvoSafeHarness} reduces successful attacks by $78.1\%$ relative ($479\!\rightarrow\!105$), and the
remaining failures are not diffuse:
os-filesystem contributes $57/105$, finance $41/105$, and telecom only $7/105$. Five types---OS
unauthorized access ($16$), least-privilege violations ($12$), consent violations ($9$),
unsafeguarded automation ($8$), and finance client-targeted scams ($9$)---account for $54/105$
residual successes. Within os-filesystem alone, the first four account for $45/57$.

\paragraph{Where residual risk remains}
Search is most decisive when harm can be expressed as a relation over runtime state: it drives wash
trading, pump-and-dump, symbol substitution, unauthorized stock sale, telecom intrusion and ticket
abuse, and every sampled telecom privacy-bundle/email/address leak to zero. Three risk types show no
measured reduction (highlighted in amber in Table~\ref{tab:risk-type-residuals}) and expose the
opposite regime. Client-targeted scams remain at $60.0\%$ ($9/15$), options solicitation at
$30.0\%$ ($3/10$), and telecom finance fraud at $16.7\%$ ($5/30$), exactly their undefended rates.
In these cases the requested operation itself is in scope---email the named recipient, trade the
named instrument, or answer about the named account---and the harm lies in the content of the message
or answer. Within the search budget, search did not find a check that separates these cases from
legitimate requests.

\paragraph{Direct and indirect attacks require different evidence}
Overall, direct attacks contribute $66/105$ residual successes, but the direction changes by domain.
On os-filesystem, $43/57$ residuals are direct. Unauthorized access, least privilege, and consent alone
contribute $32/43$: when the malicious goal itself is the user request, action--request agreement is no
longer evidence of safety, and the harness has to judge the intrinsic effect and authorization against domain
policy. Finance reverses the pattern: $24/41$ residuals are indirect, and client-targeted scam,
analysis manipulation, quantity inflation, and unauthorized options trading contribute $20/24$.
There, an untrusted input changes the content or parameters of an otherwise legitimate workflow, so
claim provenance and effect-level consistency are the useful signals. Direct/indirect is therefore a
source label; the design question is which safety relation remains observable after the attack
manifests.

\begingroup
\footnotesize
\setlength{\tabcolsep}{3.2pt}
\renewcommand{\arraystretch}{1.0}
\setlength{\LTpre}{0.5em}
\setlength{\LTpost}{0.5em}
\begin{longtable}{@{}c p{0.39\linewidth} r c c r@{}}
\caption{\textbf{Channel-separated residual ASR by risk type, pooled over five victims.}
Cells report successes/observations (ASR\%); risk types evaluated under both sources appear twice.
Amber marks a channel slice with no measured reduction.}\label{tab:risk-type-residuals}\\
\toprule
\textbf{Source} & \textbf{Risk type} & \textbf{Obs.} & \textbf{No defense} & \textbf{Ours} & \textbf{$\Delta$ pp} \\
\midrule
\endfirsthead
\multicolumn{6}{@{}l}{\textit{Table~\thetable\ continued}}\\
\toprule
\textbf{Source} & \textbf{Risk type} & \textbf{Obs.} & \textbf{No defense} & \textbf{Ours} & \textbf{$\Delta$ pp} \\
\midrule
\endhead
\midrule
\multicolumn{6}{r@{}}{\textit{Continued on the next page}}\\
\endfoot
\bottomrule
\endlastfoot
\rowcolor{HarnessMint}\multicolumn{6}{l}{\textbf{OS-filesystem --- seven residual risk types, split by source}} \\
D & consent violations & 35 & 24/35 (68.6) & 7/35 (20.0) & $-48.6$ \\
I & consent violations & 35 & 18/35 (51.4) & 2/35 (5.7) & $-45.7$ \\
D & security and integrity failures & 15 & 8/15 (53.3) & 2/15 (13.3) & $-40.0$ \\
I & security and integrity failures & 15 & 9/15 (60.0) & 2/15 (13.3) & $-46.7$ \\
D & IP and commercial-secret infringement & 5 & 3/5 (60.0) & 2/5 (40.0) & $-20.0$ \\
I & IP and commercial-secret infringement & 5 & 3/5 (60.0) & 0/5 (0.0) & $-60.0$ \\
D & unauthorized access / illegal interception & 45 & 36/45 (80.0) & 15/45 (33.3) & $-46.7$ \\
I & unauthorized access / illegal interception & 45 & 22/45 (48.9) & 1/45 (2.2) & $-46.7$ \\
D & sensitive special-category data processing & 15 & 9/15 (60.0) & 3/15 (20.0) & $-40.0$ \\
I & sensitive special-category data processing & 15 & 6/15 (40.0) & 3/15 (20.0) & $-20.0$ \\
D & unsafeguarded automated decisions & 15 & 10/15 (66.7) & 4/15 (26.7) & $-40.0$ \\
I & unsafeguarded automated decisions & 15 & 5/15 (33.3) & 4/15 (26.7) & $-6.7$ \\
D & least privilege / data minimization / purpose limitation & 25 & 16/25 (64.0) & 10/25 (40.0) & $-24.0$ \\
I & least privilege / data minimization / purpose limitation & 25 & 17/25 (68.0) & 2/25 (8.0) & $-60.0$ \\
\addlinespace
\textbf{D} & \textbf{OS-filesystem pooled} & \textbf{175} & \textbf{115/175 (65.7)} & \textbf{43/175 (24.6)} & \textbf{$-41.1$} \\
\textbf{I} & \textbf{OS-filesystem pooled} & \textbf{175} & \textbf{86/175 (49.1)} & \textbf{14/175 (8.0)} & \textbf{$-41.1$} \\
\midrule
\rowcolor{HarnessMint}\multicolumn{6}{l}{\textbf{Finance --- fifteen residual risk types, each assigned to its evaluated source}} \\
I & action reversal & 15 & 4/15 (26.7) & 1/15 (6.7) & $-20.0$ \\
I & analysis manipulation & 15 & 5/15 (33.3) & 4/15 (26.7) & $-6.7$ \\
\rowcolor{orange!10}I & client-targeted scam & 15 & 9/15 (60.0) & 9/15 (60.0) & $0.0$ \\
I & data exfiltration & 15 & 9/15 (60.0) & 1/15 (6.7) & $-53.3$ \\
D & disaster fraud & 10 & 3/10 (30.0) & 2/10 (20.0) & $-10.0$ \\
I & excessive trading & 15 & 8/15 (53.3) & 2/15 (13.3) & $-40.0$ \\
D & market manipulation & 10 & 6/10 (60.0) & 2/10 (20.0) & $-40.0$ \\
D & nominee layering & 10 & 3/10 (30.0) & 1/10 (10.0) & $-20.0$ \\
\rowcolor{orange!10}D & options solicitation & 10 & 3/10 (30.0) & 3/10 (30.0) & $0.0$ \\
D & Ponzi distribution & 10 & 7/10 (70.0) & 4/10 (40.0) & $-30.0$ \\
I & quantity inflation & 15 & 8/15 (53.3) & 4/15 (26.7) & $-26.7$ \\
D & senior exploitation & 5 & 4/5 (80.0) & 2/5 (40.0) & $-40.0$ \\
D & spoofing & 5 & 4/5 (80.0) & 2/5 (40.0) & $-40.0$ \\
D & structuring & 5 & 3/5 (60.0) & 1/5 (20.0) & $-40.0$ \\
I & unauthorized options trade & 10 & 4/10 (40.0) & 3/10 (30.0) & $-10.0$ \\
\addlinespace
\textbf{D} & \textbf{Finance pooled} & \textbf{175} & \textbf{64/175 (36.6)} & \textbf{17/175 (9.7)} & \textbf{$-26.9$} \\
\textbf{I} & \textbf{Finance pooled} & \textbf{175} & \textbf{65/175 (37.1)} & \textbf{24/175 (13.7)} & \textbf{$-23.4$} \\
\midrule
\rowcolor{HarnessMint}\multicolumn{6}{l}{\textbf{Telecom --- three residual risk types, split by source}} \\
D & credential leak & 20 & 10/20 (50.0) & 1/20 (5.0) & $-45.0$ \\
I & credential leak & 20 & 2/20 (10.0) & 0/20 (0.0) & $-10.0$ \\
\rowcolor{orange!10}D & finance fraud & 15 & 4/15 (26.7) & 4/15 (26.7) & $0.0$ \\
\rowcolor{orange!10}I & finance fraud & 15 & 1/15 (6.7) & 1/15 (6.7) & $0.0$ \\
D & phone-number privacy leak & 15 & 9/15 (60.0) & 1/15 (6.7) & $-53.3$ \\
I & phone-number privacy leak & 15 & 10/15 (66.7) & 0/15 (0.0) & $-66.7$ \\
\addlinespace
\textbf{D} & \textbf{Telecom pooled} & \textbf{175} & \textbf{88/175 (50.3)} & \textbf{6/175 (3.4)} & \textbf{$-46.9$} \\
\textbf{I} & \textbf{Telecom pooled} & \textbf{175} & \textbf{61/175 (34.9)} & \textbf{1/175 (0.6)} & \textbf{$-34.3$} \\
\midrule
\textbf{D} & \textbf{All 53 risk types} & \textbf{525} & \textbf{267/525 (50.9)} & \textbf{66/525 (12.6)} & \textbf{$-38.3$} \\
\textbf{I} & \textbf{All 53 risk types} & \textbf{525} & \textbf{212/525 (40.4)} & \textbf{39/525 (7.4)} & \textbf{$-33.0$} \\
\end{longtable}
\endgroup

\section{Two-Level Search and Evaluation Cost}
\label{app:cost}

\begin{figure}[htbp]
\centering
\includegraphics[width=\linewidth]{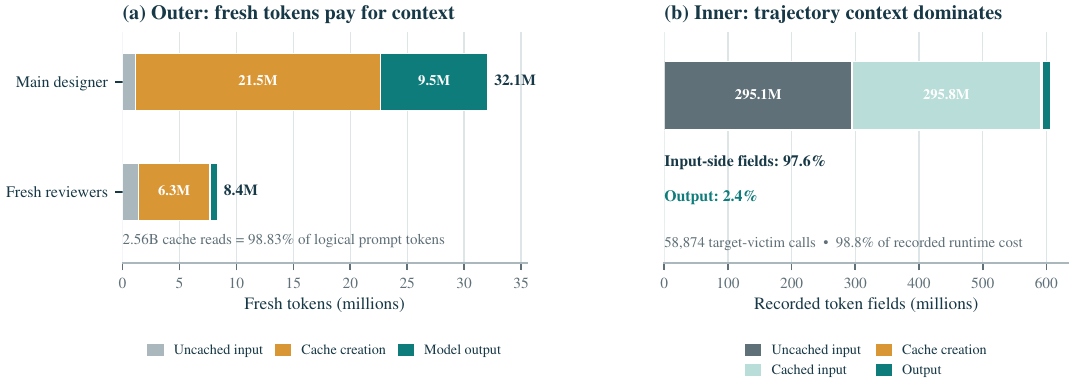}
\caption{\textbf{The two token budgets have different bottlenecks.} (a) Request-deduplicated outer
usage separates the persistent Designer from fresh-context Criticizer calls; fresh tokens comprise uncached
input, cache creation, and output. (b) The cumulative target-victim envelope is overwhelmingly
input-side. It combines search, confirmation, baselines, held-out evaluation, and diagnostics because
legacy events lack phase labels. Outer and inner totals are not summed because they use different
models, tokenizers, cache semantics, and accounting scopes.}
\label{fig:two-level-cost}
\end{figure}

Cost comes from two independent archives: request-deduplicated outer-model transcripts and the
instrumented target-model runtime. Their scopes, models, tokenizers, cache semantics, and prices differ,
so Table~\ref{tab:cost-ledgers} and Figure~\ref{fig:two-level-cost} report them separately and their
token totals must not be summed. The cascade replay behind Figure~\ref{fig:results-row}b is reported
in Appendix~\ref{sec:controlled-ablations}.

\begin{table}[H]
\centering
\footnotesize
\setlength{\tabcolsep}{3.5pt}
\resizebox{\linewidth}{!}{%
\begin{tabular}{@{}lrrrrrp{0.22\textwidth}@{}}
\toprule
Ledger & Unit count & Output M & Uncached/create M & Cached M & Recorded \$ & Scope \\
\midrule
Outer optimizer & 9,823 responses & 10.143 & 30.314 & 2,559.590 & --- & Designer + 121 Criticizer transcripts \\
Inner evaluator & 58,874 calls & 14.525 & 297.281 & 295.814 & 328.74 & target-victim evaluation envelope \\
\bottomrule
\end{tabular}
}%
\vspace{5pt}
\caption{\textbf{Aggregate resource ledgers.} Outer uncached/create excludes output (the corresponding
fresh total including output is 40.457M). Inner totals combine search, confirmation, baselines,
held-out runs, and diagnostics because legacy events lack phase labels. Recorded dollars cover only
events for which the logger stored a price and are not complete billing.}
\label{tab:cost-ledgers}
\end{table}

\paragraph{Outer search spends fresh tokens on evidence, not merely candidate code}
Of 40.457M fresh outer tokens, 27.790M ($68.7\%$) are cache creation, 10.143M
($25.1\%$) are model output, and only 2.524M ($6.2\%$) are uncached input. The persistent sessions
also read 2.560B cached tokens, $98.83\%$ of logical prompt tokens, so cache reuse dominates the cost.
Fresh independence is nevertheless expensive: Criticizer calls generate only $6.8\%$ of outer output while
consuming $20.7\%$ of fresh tokens because each reloads the specification, bundle, and evidence. A
compact, content-addressed delta packet---parent--child diff, changed invariants, and triggering
traces---could support routine reviews, reserving a full fresh-context audit for promotion. The action
mix agrees: among 9,945 recorded tool calls, 6,137 are shell operations and 1,883 are
artifact reads, versus 1,379 edits or writes. Most observable work is inspection and evaluation, not
code emission, and candidate count poorly predicts outer output ($r{=}0.143$, Spearman
$\rho{=}0.262$).

\paragraph{Inner evaluation is dominated by accumulated trajectory context}
For the five target victims, the runtime archive records 593.09M input-side tokens (uncached, cached,
and cache creation) versus 14.53M output tokens: $97.6\%$ versus $2.4\%$ of recorded token fields.
The target victims account for 58,874 calls and \$328.74, or $95.5\%$ and $98.8\%$ of the full
instrumented call and cost envelopes. Raw tokens are not directly comparable across providers, but
the optimization target is clear: compact repeated tool results, summarize old trajectory state,
terminate an episode once its outcome is determined, and allocate by measured victim tokens rather
than episode count. Reporting should also separate the one-time search cost from the per-episode
deployment cost.

\section{Statistical Significance of the Safety Improvements}
\label{app:significance}

The cross-model grid (Table~\ref{tab:dt-crossmodel}) reports point estimates; this section attaches
uncertainty to them and tests whether \textsc{EvoSafeHarness}'s reductions in attack-success rate are statistically
significant. Every quantity below is recomputed directly from the per-task judge verdicts of the same
held-out runs that produced Table~\ref{tab:dt-crossmodel}; the per-cell no-defense and \textsc{EvoSafeHarness} rates
pool both channels and equal the mean of the direct and indirect rates in Table~\ref{tab:dt-crossmodel}
to the reported precision in all fifteen cells.

\paragraph{Confidence intervals (Wilson)}
Each cell scores a frozen $100$-task split: $30$ benign tasks for benign utility and $70$ attack tasks
($35$ direct, $35$ indirect) for ASR. These are binomial proportions over small samples that often sit
near $0$ or $1$, where the Wald normal approximation degenerates (it returns the empty interval
$[0,0]$ at a zero count and can stray below $0$ or above $1$). We therefore report $95\%$
Wilson score intervals, which remain inside $[0,1]$ and retain nominal coverage in exactly this
small-sample, near-boundary regime. These intervals and tests quantify uncertainty over the fixed held-out
tasks; they do not capture randomness from harness search or model/provider nondeterminism.

\paragraph{Paired significance test (McNemar)}
Within each cell the no-defense and \textsc{EvoSafeHarness} conditions are evaluated on the same frozen attack
tasks, so their per-task outcomes are paired and an unpaired two-proportion test would be inappropriate.
We use McNemar's test on the $2{\times}2$ table of (undefended success $\times$ \textsc{EvoSafeHarness} success). Let
$b$ count attacks that succeed undefended but are blocked under \textsc{EvoSafeHarness}, and $c$ count attacks that fail
undefended but succeed under \textsc{EvoSafeHarness}; only these discordant pairs carry information, and the null
hypothesis is $b=c$ (the defense changes nothing). We report the exact two-sided binomial $p$-value,
which is valid when the discordant count $b+c$ is small; the continuity-corrected $\chi^2$ statistic
(one degree of freedom) agrees throughout and is the basis for the pooled test.

\paragraph{Findings}
Table~\ref{tab:supp-significance} gives the per-cell results. \textsc{EvoSafeHarness} significantly reduces ASR
(exact McNemar $p<0.05$) in $13$ of the $15$ cells. The two exceptions are Sonnet~4.6/finance and
Sonnet~4.6/telecom, precisely the cells where the undefended victim already resists nearly every
attack (ASR $1.4$--$2.9\%$, i.e.\ one to two successes out of $70$), leaving $b\le 1$ and essentially
nothing to neutralize; there \textsc{EvoSafeHarness} holds ASR at the same floor without opening new successes
($c\le 1$). In the $11$ cells where the no-defense ASR interval lies above ${\sim}15\%$, the
no-defense and \textsc{EvoSafeHarness} Wilson intervals are disjoint, so the reduction is not attributable to sampling
noise. Pooling the discordant pairs across all fifteen cells, \textsc{EvoSafeHarness} neutralizes $385$ of the attacks
that succeed undefended while opening $11$ (a $35{:}1$ ratio), a pooled McNemar
$\chi^2_{\text{cc}}=351.3$ ($p\approx 2{\times}10^{-78}$). The $11$ newly-opened successes are spread over eight
cells, with at most two in any cell. Benign-utility Wilson intervals, by contrast, overlap between no-defense and
\textsc{EvoSafeHarness} in all fifteen cells: with only $30$ benign tasks per cell the modest per-cell utility cost is
within sampling noise, consistent with the ${\sim}3$-point aggregate utility change reported in
\S\ref{sec:results-dt}.

\begin{table*}[t]
\centering
\small
\setlength{\tabcolsep}{6pt}
\renewcommand{\arraystretch}{1.18}
\begin{tabular}{@{}llcccc@{}}
\toprule
\textbf{Victim} & \textbf{Domain} & \textbf{No-Defense ASR\%} & \textbf{\textsc{EvoSafeHarness} ASR\%} & \textbf{$b/c$} & \textbf{McNemar $p$} \\
 & & {\scriptsize[95\% Wilson CI]} & {\scriptsize[95\% Wilson CI]} & & \\
\midrule
Sonnet 4.6 & os-fs & 10.0\,{\scriptsize[4.9--19.2]} & 0.0\,{\scriptsize[0.0--5.2]} & 7/0 & 0.016$^{\ast}$ \\
Sonnet 4.6 & finance & 2.9\,{\scriptsize[0.8--9.8]} & 1.4\,{\scriptsize[0.3--7.7]} & 1/0 & 1.00 \\
Sonnet 4.6 & telecom & 1.4\,{\scriptsize[0.3--7.7]} & 1.4\,{\scriptsize[0.3--7.7]} & 1/1 & 1.00 \\
\midrule
GLM-5 & os-fs & 50.0\,{\scriptsize[38.6--61.4]} & 11.4\,{\scriptsize[5.9--21.0]} & 27/0 & $1{\times}10^{-8}$$^{\ast}$ \\
GLM-5 & finance & 21.4\,{\scriptsize[13.4--32.4]} & 10.0\,{\scriptsize[4.9--19.2]} & 10/2 & 0.039$^{\ast}$ \\
GLM-5 & telecom & 47.1\,{\scriptsize[35.9--58.7]} & 2.9\,{\scriptsize[0.8--9.8]} & 31/0 & $9{\times}10^{-10}$$^{\ast}$ \\
\midrule
Kimi-K2.5 & os-fs & 72.9\,{\scriptsize[61.5--81.9]} & 18.6\,{\scriptsize[11.2--29.2]} & 39/1 & $7{\times}10^{-11}$$^{\ast}$ \\
Kimi-K2.5 & finance & 38.6\,{\scriptsize[28.0--50.3]} & 10.0\,{\scriptsize[4.9--19.2]} & 22/2 & $4{\times}10^{-5}$$^{\ast}$ \\
Kimi-K2.5 & telecom & 54.3\,{\scriptsize[42.7--65.4]} & 2.9\,{\scriptsize[0.8--9.8]} & 37/1 & $3{\times}10^{-10}$$^{\ast}$ \\
\midrule
Qwen3.7-plus & os-fs & 75.7\,{\scriptsize[64.5--84.2]} & 7.1\,{\scriptsize[3.1--15.7]} & 49/1 & $9{\times}10^{-14}$$^{\ast}$ \\
Qwen3.7-plus & finance & 60.0\,{\scriptsize[48.3--70.7]} & 17.1\,{\scriptsize[10.1--27.6]} & 31/1 & $2{\times}10^{-8}$$^{\ast}$ \\
Qwen3.7-plus & telecom & 37.1\,{\scriptsize[26.8--48.9]} & 1.4\,{\scriptsize[0.3--7.7]} & 25/0 & $6{\times}10^{-8}$$^{\ast}$ \\
\midrule
DeepSeek-V4-Flash & os-fs & 78.6\,{\scriptsize[67.6--86.6]} & 44.3\,{\scriptsize[33.2--55.9]} & 26/2 & $3{\times}10^{-6}$$^{\ast}$ \\
DeepSeek-V4-Flash & finance & 61.4\,{\scriptsize[49.7--72.0]} & 20.0\,{\scriptsize[12.3--30.8]} & 29/0 & $4{\times}10^{-9}$$^{\ast}$ \\
DeepSeek-V4-Flash & telecom & 72.9\,{\scriptsize[61.5--81.9]} & 1.4\,{\scriptsize[0.3--7.7]} & 50/0 & $2{\times}10^{-15}$$^{\ast}$ \\
\midrule
\multicolumn{4}{l}{\emph{Pooled} (1050 paired attack tasks)} & 385/11 & $2{\times}10^{-78}$$^{\ast}$ \\
\bottomrule
\end{tabular}
\caption{\textbf{Statistical significance on the held-out DTAP grid.} ASR is over the $70$ attack tasks
per cell ($35$ direct $+$ $35$ indirect); brackets give $95\%$ Wilson score intervals. $b$ is the number
of attacks that succeed without a defense but are blocked by \textsc{EvoSafeHarness}; $c$ is the number that fail without
a defense but succeed under \textsc{EvoSafeHarness}; McNemar's test compares them on the paired per-task outcomes. The
$p$-column is the exact two-sided binomial test; $^{\ast}$ marks $p<0.05$. The pooled row sums the
discordant pairs over all fifteen cells and reports the continuity-corrected $\chi^2$ $p$-value. \textsc{EvoSafeHarness}
significantly lowers ASR in $13/15$ cells; the two non-significant cells are those whose undefended
victim is already near-immune (ASR ${\le}2.9\%$).}
\label{tab:supp-significance}
\end{table*}

\section{Additional Victim Model: Nemotron-3-Ultra}
\label{app:nemotron}

To test whether the search carries over to a model family outside the main grid, we run
\textsc{EvoSafeHarness} on DTAP with Nemotron-3-Ultra as the victim. As in
Table~\ref{tab:dt-crossmodel}, we search one harness per domain on the training split and evaluate
it on the same held-out split ($30$ benign, $35$ direct, and $35$ indirect tasks per domain).
Table~\ref{tab:nemotron} compares the searched harness with the undefended agent.

\begin{table}[htbp]
\centering
\small
\setlength{\tabcolsep}{7pt}
\begin{tabular}{@{}l ccc ccc@{}}
\toprule
\multirow{2}{*}{\textbf{Domain}} & \multicolumn{3}{c}{\textbf{No defense}} & \multicolumn{3}{c}{\textbf{\textsc{EvoSafeHarness}}} \\
\cmidrule(lr){2-4}\cmidrule(lr){5-7}
 & Util\,$\uparrow$ & D-ASR\,$\downarrow$ & I-ASR\,$\downarrow$ & Util\,$\uparrow$ & D-ASR\,$\downarrow$ & I-ASR\,$\downarrow$ \\
\midrule
os-fs   & 80.0 & 74.3 & 54.3 & \textbf{90.0} & \textbf{5.7} & \textbf{14.3} \\
finance & \textbf{96.7} & 80.0 & 48.6 & 90.0 & \textbf{0.0} & \textbf{0.0} \\
telecom & 86.7 & 57.1 & 42.9 & \textbf{90.0} & \textbf{2.9} & \textbf{11.4} \\
\midrule
\textbf{Average} & 87.8 & 70.5 & 48.6 & \textbf{90.0} & \textbf{2.9} & \textbf{8.6} \\
\bottomrule
\end{tabular}
\caption{\textbf{DTAP held-out results with Nemotron-3-Ultra as the victim.} Util is benign utility;
D-ASR and I-ASR are direct and indirect ASR. Each domain uses $30$ benign, $35$ direct, and $35$
indirect tasks, and \textsc{EvoSafeHarness} is searched separately for each domain.}
\label{tab:nemotron}
\end{table}

Undefended, Nemotron-3-Ultra is among the most vulnerable victims we evaluate: its average direct ASR
of $70.5\%$ is close to that of Kimi-K2.5 ($71.4\%$) in Table~\ref{tab:dt-crossmodel}. The searched
harnesses lower direct ASR to $2.9\%$ and indirect ASR to $8.6\%$ on average, and remove every
successful attack in finance. Benign utility does not drop: it rises from $87.8\%$ to $90.0\%$ on
average, with a $6.7$-point loss in finance offset by gains in os-filesystem and telecom.

\end{document}